%% file: paper.tex
\documentclass[sigconf]{acmart}

\setcopyright{none}
\renewcommand\footnotetextcopyrightpermission[1]{}

\copyrightyear{2025} 
\acmYear{2025} 
\acmConference[]{Preprint}{2025}{ }
\acmBooktitle{ACM eEnergy}

\usepackage{tabularx,adjustbox}
\usepackage{graphicx}  
\usepackage{algorithm,algpseudocode} 
\usepackage{longfbox}
\usepackage{enumitem}

\usepackage{amsmath,amsfonts,mathtools,mathrsfs}

\usepackage[sans]{dsfont}

\usepackage{array}
\usepackage{url}

\makeatletter
\newsavebox{\@brx}
\newcommand{\llangle}[1][]{\savebox{\@brx}{\(\m@th{#1\langle}\)}%
  \mathopen{\copy\@brx\kern-0.5\wd\@brx\usebox{\@brx}}}
\newcommand{\rrangle}[1][]{\savebox{\@brx}{\(\m@th{#1\rangle}\)}%
  \mathclose{\copy\@brx\kern-0.5\wd\@brx\usebox{\@brx}}}
\makeatother

\newcommand{\raf}[1]{(\ref{#1})}

\newtheorem{innercustomthm}{\textbf{Theorem}}
\newenvironment{customthm}[1]
{\renewcommand\theinnercustomthm{#1}\innercustomthm}
{\endinnercustomthm}

\algnewcommand{\LeftComment}[1]{\Statex \(\triangleright\) #1}

\usepackage{subcaption}
\begin{document}

\title[Blockchain-Enabled Decentralized Privacy-Preserving Group Purchasing for Energy Plans]{Blockchain-Enabled Decentralized Privacy-Preserving \\ Group Purchasing for Retail Energy Plans}

\author{Sid Chi-Kin Chau}\authornote{Corresponding author: \url{sid.chau@acm.org}}
\orcid{0000-0003-0362-2844}
\affiliation{\institution{CSIRO Data61}
\city{Sydney}
\country{Australia}}
\email{sid.chau@acm.org}

\author{Yue Zhou}
\email{yue.zhou@anu.edu.au}
\affiliation{\institution{Australian National University}
\city{Canberra}
\country{Australia}}

\renewcommand{\shortauthors}{S. C.-K. Chau, Y. Zhou}

\makeatletter
\makeatother

\begin{abstract}
\input{abstract}
\end{abstract}

% remove ACM Reference after Abstract
%\settopmatter{printacmref=false}



\settopmatter{printfolios=true, printacmref=false} 
\maketitle

\input{intro}

\input{model}

\input{model2}

\input{spdz}

\input{algo}

\input{experiment}

\input{related}

\input{concl}

\bibliographystyle{ACM-Reference-Format}
\bibliography{reference}

\appendix
\section*{Appendix}

\input{append1}

\input{append3}

\input{append0}

\end{document}

%% file: abstract.tex
Retail energy markets are increasingly consumer-oriented, thanks to a growing number of energy plans offered by a plethora of energy suppliers, retailers and intermediaries. To maximize the benefits of competitive retail energy markets, {\em group purchasing} is an emerging paradigm that aggregates consumers' purchasing power by coordinating switch decisions to specific energy providers for discounted energy plans. Traditionally, group purchasing is mediated by a trusted third-party, which suffers from the lack of privacy and transparency. In this paper, we introduce a novel paradigm of {\em decentralized privacy-preserving} group purchasing, empowered by privacy-preserving blockchain and secure multi-party computation, to enable users to form a coalition for coordinated switch decisions in a decentralized manner, without a trusted third-party. The coordinated switch decisions are determined by a competitive online algorithm, based on users' private consumption data and current energy plan tariffs. Remarkably, no private user consumption data will be revealed to others in the online decision-making process, which is carried out in a transparently verifiable manner to eliminate frauds from dishonest users and supports fair mutual compensations by sharing the switching costs to incentivize group purchasing. We implemented our decentralized group purchasing solution as a smart contract on Solidity-supported blockchain platform (e.g., Ethereum), and provide extensive empirical evaluation.\footnote{This is an extended version of the conference paper \cite{CZ22energyplan} appearing in ACM International Conference on Future and Sustainable Energy Systems (ACM e-Energy 2022)}.\enlargethispage{15pt}

%% file: intro.tex
\section{Introduction}

With rising deregulation and decarbonization in the traditionally monopolized energy sector, a myriad of energy plans and tariffs are being offered in retail energy markets in the US, Europe, and other countries \cite{zhou17retail}. In today's competitive retail energy markets with increasing choices, users are able to compare and switch among diverse energy plans from multiple retail energy providers \cite{ZCC19energyplan}, with different tariff structures (e.g., hourly rates, peak/off-peak hours, PV feed-in tariffs) and contractual arrangements (e.g., connection/disconnection fees, contracted periods). Some energy plans offer special subsidies (e.g., incentive packages for home batteries and energy-efficient boilers), and options for renewable energy and carbon offsetting. With a growing set of energy suppliers and retailers, the number of energy plans has mushroomed significantly. For instance, 160+ energy plans are available in Buffalo, the US \cite{eastcoast}, while 400+ electricity plans are available in Sydney, Australia \cite{energymadeeasy}. This presents ample opportunities for end users to cherry-pick the best plans that optimize their energy bills and needs.

Along with a variety of energy plans, a new consumer-driven paradigm called {\em group purchasing} (or bulk buying) has emerged in retail energy markets \cite{ZC21energyplan}, which boosts consumers' purchasing power by coordinated purchasing decisions in a coalition of consumers. There has been a long history of group purchasing in various sectors \cite{CR11group}. The idea that consumers should aggregate their demands to increase bargaining power with vendors has been practiced in health care and e-commerce (e.g., Groupons, Meituan). Recently, there emerged several group purchasing start-ups in the energy sector that operate as intermediaries between energy suppliers and users. Some of them recruit users for discounted group-based energy plans, whereas others solicit group-based tenders from energy suppliers on behalf of their users (e.g., ChoiceEnergy \cite{choiceenergy}). With group purchasing, users can collectively switch to specific energy suppliers for better discounts by leveraging their aggregate purchasing power. Hence, group purchasing for energy plans is becoming a popular paradigm in retail energy markets.

Traditionally, group purchasing is mediated by a trusted third-party agent in a centralized manner, such that users are required to submit their personal profiles and private data (e.g., past energy bills) to an agent, who will negotiate with some vendors on their behalf. In exchange for discounted energy plans for users, the agent will receive commission fees from the vendors. However, there is a lack of {\em transparency} in the third-party mediation approach. The agent may obscure the commission fees and negotiation process, and may not maximize the users' interest, but rather the interest of the agent or the vendors. Moreover, there is a {\em privacy} concern -- users may be unaware of the potential misuses and breaches of their private data by the agent for other unintended purposes. To bolster user privacy, stricter privacy protection legislations (e.g., GDPR in Europe) are introduced in various countries to restrict personal information access by a third-party. Yet, private user data can still be exploited through other illicit approaches (e.g., hacking into the agent).  Therefore, there is a need to ensure both transparency and privacy in the decision-making processes of group purchasing, while enjoying its benefits.

\begin{figure*}[bht!] 
\includegraphics[width=\textwidth]{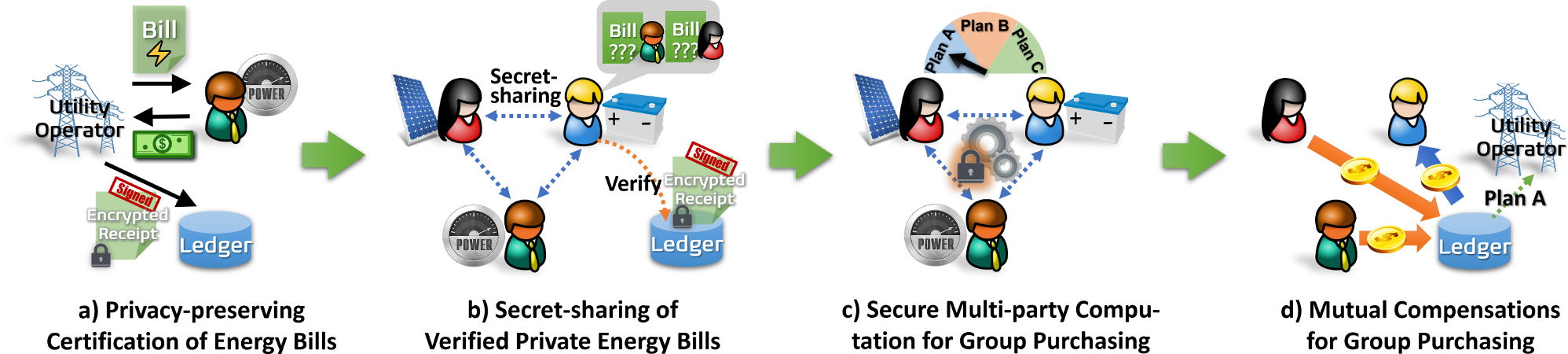} 
\caption{An illustration of our solution for decentralized privacy-preserving group purchasing.}
\label{fig:scenario}
\end{figure*}

\smallskip

In this paper, we introduce a novel paradigm of ``{\em decentralized privacy-preserving group purchasing}'', which removes the trusted third-party intermediaries. Our goal is to enable users to form a coalition for making coordinated energy plan switch decisions in a decentralized manner with assurance of privacy and transparency: 

\begin{enumerate}

\item {\bf Privacy Protection of User Data:} 
Deciding a suitable energy plan requires a sophisticated consideration of a variety of factors, such as past consumption data and current energy plan tariffs. However, sharing personal information with others in group purchasing may be undesirable. We need to guarantee that no private user consumption data will be revealed to other parties for unintended purposes, while ensuring a discreet decision-making process that properly incorporates the data of all users in group purchasing. 

\item {\bf Transparently Verifiable Decision-Making:}
Without the knowledge of its private input, privacy poses a significant challenge to the verifiability of the decision-making process. In particular, dishonest users may take advantage of privacy protection to cheat or misrepresent their energy data. These dishonest users are likely to collude to coordinate their actions. Hence, it is critical to safeguard against the potential presence of a large number of dishonest users. We need to ensure that the decision-making process should be transparent and verifiable to eliminate any fraud from dishonest users, while preserving user privacy.

\end{enumerate}

In this paper, we propose an effective solution to support decentralized privacy-preserving group purchasing. Our solution draws on several components: (1)  a competitive online algorithm for group purchasing decision-making, (2) secure multi-party computation for privacy-preserving online decision-making, and (3) zero-knowledge proofs on blockchain for validating the private input data to our online algorithm. We briefly explain these components as follows, but the details will be elaborated in the subsequent sections.

\smallskip

{\bf\em Competitive Online Algorithm:}
First, the problem of energy plan selection \cite{ZCC19energyplan} without future knowledge (e.g., future energy demands and tariffs of future energy plans) belongs to {\em online decision-making problems}. There is an extensive body of literature \cite{BEY05online} on online algorithms that solve these online problems with theoretical-proven bounds on the optimality of their online decisions (as known as competitive ratios). In this paper, we formulate an online decision problem of group purchasing for energy plan as a Metrical Task System problem, and devise a competitive online algorithm that produces close-to-optimal performance in our evaluation. Our online algorithm extends the related work \cite{ZC21energyplan}. Furthermore, our online algorithm supports {\em mutual compensations} in group purchasing, such that some users may be fairly compensated by others for the switching costs to join group purchasing. Mutual compensations effectively incentivize group purchasing. 

{\bf\em Secure Multi-party Computation:} 
Second, we execute our competitive online algorithm for group purchasing in a privacy-preserving manner, without disclosing the private input data (e.g., users' energy bills). We rely on secure multi-party computation (or simply called multi-party computation (MPC)), which is a general framework to compute a function jointly by multiple parties with concealed input from each other. Recently, SPDZ \cite{cramer2015secure}, an efficient MPC protocol based on secret-sharing, has been applied to many practical applications (e.g., machine learning \cite{cpr18spdz}). SPDZ can safeguard against a dishonest majority of users ($\ge 50\%$ of users may be dishonest). In this paper, we apply SPDZ to the online decision-making process of group purchasing with concealed input data. 

 {\bf\em Blockchain and Zero-knowledge Proofs:}
Third, it is important to validate the private input data, before the online decision-making process. Otherwise, dishonest users may input falsified data to influence the outcomes for their benefits.  We utilize blockchain to record users' energy bills, and employ smart contracts for verifying mutual compensations among users. There are several functions of blockchain. (i) Blockchain is a decentralized platform for verifiable applications without trusted intermediaries, which enables decentralized verification of mutual compensations to match the decision of the online algorithm. (ii) The crypto-tokens on blockchain serve as a convenient micro-payment system for compensations among users. (iii) Blockchain is a public non-temperable data repository for energy bill data. 
However, blockchain by default does not protect privacy, and the data is disclosed publicly on the ledger. Therefore, we apply zero-knowledge proofs (ZKPs), which can verify certain properties of encrypted data, without decrypting it. For example, instead of storing users' energy bills, we record the corresponding encrypted receipts on blockchain. Then users can use ZKPs to validate the input data for online decision-making with respect to the encrypted receipts, without disclosing the original data.

\smallskip

Together with these components, we develop an effective solution to support decentralized privacy-preserving group purchasing, as illustrated at a high level in Figure~\ref{fig:scenario}:
\begin{enumerate}

\item[a)] First, the utility operator records the encrypted receipts of energy bills on the blockchain ledger, after users pay their bills. But no private energy bills are revealed on the ledger.

\item[b)] The users use multi-party computation (SPDZ) to share their energy bill data in a privacy-preserving manner, without disclosing the private data to each other. They will jointly verify the validity of private data, via SPDZ, based on ZKPs.

\item[c)] With verified private input data, the users jointly execute the online algorithm via SPDZ to decide the suitable group purchasing decision that maximizes the benefits of all users, if possible. Moreover, they will also decide mutual compensations to incentivize each other to join group purchasing. 

\item[d)] The group purchasing decision and mutual compensations are handled and recorded on the blockchain. The utility operator executes the chosen energy plan according to the records on the blockchain. The users are assured of transparency in the whole decision-making process.

\end{enumerate}

The technical details of the procedures will be presented in the rest of the paper. We remark that although our solution is designed for energy plan selection, it is rather generic and generalizable to other group purchasing applications (e.g., health care plans).

\smallskip

This paper is organized as follows. We first formulate a decision-making model, in which users form a coalition for group purchasing of energy plans without the knowledge of future energy demands and tariffs. In particular, we consider mutual compensations in online decision-making to incentivize group purchasing. We next formulate the security model considering privacy protection and possible attacks from dishonest users. We then present the basics of cryptographic components and multi-party computation. The privacy-preserving solution is presented with an evaluation of our implementation on Solidity-supported blockchain platform.

%% file: model.tex
\section{Decision-Making Model}  \label{sec:model}

In the following, we formulate a problem of group purchasing for energy plans as an online decision-making problem that is possibly conducted in a centralized manner, without considering privacy. In the subsequent sections, we will incorporate privacy protection in a decentralized mechanism. First, we will consider the standalone setting of energy plan selection without group purchasing. Then we will extend our study to the setting with group purchasing. Finally, we will present a group purchasing decision-making mechanism.

\subsection{Problem Setup}

Our problem of energy plan selection is consisted of the following components:
\begin{itemize}

\item {\bf Energy Plans:}
We consider discrete timeslots, indexed by $t \in \{1,...,T\}$. There are a set of energy plans ${\mathcal E}$. Each energy plan is indexed by $j \in {\mathcal E}$ and characterized by a tuple ${\tt E}(j) = \langle {\tt p}^t_+(j), {\tt p}^t_{\mbox{-}}(j), {\tt c}(j), {\tt d}(j),  {\tt T}(j)\rangle$, as described as follows:

\begin{enumerate}

\item {\em Consumption Tariffs}: Let ${\tt p}^t_{+}(j)$ be a pricing function that maps a timeslot $t$ to its respective time-of-use price of energy consumption (i.e., energy import). ${\tt p}^t_{+}(j)$ can capture a different rate at timeslot $t$ for the peak, off-peak and shoulder periods. 

\item {\em Feed-in Tariffs}: We also consider the feed-in rates of energy production (i.e., energy export from PV or home batteries). Let ${\tt p}^t_{\mbox{-}}(j)$ be a pricing function that maps a timeslot $t$ to its respective time-varying price of energy production. 

\item {\em Connection Fee}: Let ${\tt c}(j)$ be the connection fee when a user joins energy plan $j$. Typical connection fee includes the set-up and installation costs at the distribution grid.

\item {\em Disconnection Fee}: Let ${\tt d}(j)$ be the disconnection fee when a user terminates energy plan $j$ before the contract period. Higher disconnection fees can discourage users from early termination of the energy plans. For simplicity, we consider constant disconnection fees, regardless the residual period of the contract duration.

\item {\em Contract Duration}:  Let ${\tt T}(j)$ be the contract duration, beyond which the energy plan can be terminated without incurring any disconnection fee. %We assume that ${\tt T}(j)$ is sufficiently long that a user needs to explicitly consider the disconnection fee when switching to another energy plan.	

\end{enumerate}

\item {\bf Energy Users:}  There are a set of users ${\mathcal N}$, indexed by $i \in \{1, ..., N\}$. Each user $i$ has certain energy consumption or production rates over time, represented by a sequence $(a_i^t)$, where $a_i^t > 0$ denotes energy consumption and $a_i^t < 0$ denotes energy production. Let $x_i^t \in {\mathcal E}$ be the selected energy plan by user $i$ at timeslot $t$. Let $x_i = (x_i^1,...,x_i^T)$ be a sequence of user $i$'s selections over time, and its feasible space be ${\mathcal E}^T$. 

Let the long-term energy cost incurred by selections $x_i$ be
\begin{align} 
{\tt Cost}_i[x_{i}] \triangleq & \sum_{t=1}^T \Big( {\tt Cost}_{\tt Op}^t[{x_i^t},a_i^t] + {\tt Cost}_{\tt Sw}^t[{x_i^{t-1}},{x_i^t}]\Big)
\end{align} 
where ${\tt Cost}_{\tt Op}^t[{x_i^t},a_i^t]$ is the operational cost defined by
$${\tt Cost}_{\tt Op}^t[{x_i^t},a_i^t] \triangleq {\tt p}_+^t(x_i^t) \cdot a_i^t\cdot {\mathds 1}\{a_i^t\ge 0\} + {\tt p}_{\mbox{-}}^t(x_i^t) \cdot a_i^t\cdot {\mathds 1}\{a_i^t< 0\}$$
and ${\tt Cost}_{\tt Sw}^t[{x_i^{t-1}},{x_i^t}]$ is the switching cost between energy plans defined by
$${\tt Cost}_{\tt Sw}^t[{x_i^{t-1}},{x_i^t}] \triangleq  \Big({\tt c}({x_i^t}) + {\tt d}({x_i^{t-1}}) \cdot {\mathds 1}\{{\tt T}_i^t \le {\tt T}({x_i^{t-1}})\} \Big) \cdot {\mathds 1}\{x_i^t \ne x_i^{t-1}\}$$
Let ${\tt T}_i^t$ be the number of previous consecutive timeslots that user $i$ has been using energy plan $x_i^{t-1}$ before timeslot $t$.

The goal of each user $i$ is to decide $x_i$, as to minimize her long-term total energy cost in the following problem:
\begin{equation} \label{eqn:mincost}
\begin{array}{lc}
 ({\sf P}_i) \ \ \ & \min_{x_i \in {\mathcal E}^T} {\tt Cost}_i[x_{i}]
 \end{array}
\end{equation}
However, each user can only do so in an online manner without the future knowledge of $a_i^{t'}$, ${\tt p}^{t'}_{+}(j)$ and ${\tt p}^{t'}_{\mbox{-}}(j)$ for any $j$ and $t' > t$ at each timeslot $t$, because there may be uncertain future demands and tariffs, or new energy plans.
%For simplicity, we assume that ${\tt T}(j)$ is sufficiently long, and we do not consider ${\tt T}(j)$ in the switching cost $({\tt c}_{x_i^t} + {\tt d}_{x_i^{t-1}})$. But it is straightforward to incorporate the contract period, if needed.

\end{itemize}

\subsection{Standalone Online Energy Plan Selection} 

In this section, we solve $({\sf P}_i)$ in the standalone setting without group purchasing. This problem has been studied previously in \cite{ZCC19energyplan}.  In this subsection, we present some preliminaries for this problem.

Problem $({\sf P}_i)$ belongs to a general problem class called {\em Metrical Task System} (MTS) problem \cite{BEY05online}, which consists of a set of states (e.g., energy plans). Each state is associated with a time-varying cost, and there is a switching cost if the decision-maker switches from one state to another. The goal of the decision-maker is to decide a sequence of states to minimize the total of state costs and switching costs, without knowing the future state costs. The MTS problem captures a wide range of scenarios, such as $k$-server, energy generation dispatching, and energy plan selection problems. 

We describe the offline and online algorithms for solving $({\sf P}_i)$:

\begin{enumerate}

\item {\bf Offline Algorithm:} The offline optimal solution of $({\sf P}_i)$ can be computed by dynamic programming. Formally, let ${\tt Opt}_i^t[x_i^t]$ be the optimal cost of user $i$ with selection $x_i^t \in {\mathcal E}$ at timeslot $t$. When ${\tt Opt}_i^{t-1}[x]$ is known for all $x \in {\mathcal E}$, ${\tt Opt}^t_i[x_i^t]$ can be computed iteratively by Bellman-Ford equation:
\begin{align}
{\tt Opt}_i^t[x_i^t] = & \min_{x\in {\mathcal E}}\Big( {\tt Opt}_i^{t-1}[x] + 
{\tt Cost}_{\tt Op}^t[x_i^t,a_i^t] + {\tt Cost}_{\tt Sw}^t[x,x_i^t] \Big) \notag
%{\tt p}^t[{x_i^t},a_i^t]  + ({\tt c}_{x_i^t} +  {\tt d}_{x} )\cdot {\mathds 1}_{x_i^t \ne x} \Big)
\end{align}
Initially, we set ${\tt Opt}_i^0[x] = 0$ for all $x$. Then, we can compute ${\tt Opt}_i^t[x]$ for all $x$ using ${\tt Opt}_i^{t-1}[x]$ by dynamic programming.
Let $\big({x^\ast}^t_i\big)_{t=1}^T$ be an offline optimal selection of $({\sf P}_i)$, which is obtained by backward computation. First, we obtain ${x^\ast}^{T}_i$ by
\begin{equation}
{x^\ast}^{T}_i = \arg\min_{x\in {\mathcal E}} {\tt Opt}_i^T[x] 
\end{equation}
Then, we can obtain ${x^\ast}^{t-1}_i = x$ using ${x^\ast}^t_i$ by finding a suitable $x$ that satisfies
\begin{align}
{\tt Opt}_i^t[{x^\ast}^{t}_i] = & {\tt Opt}_i^{t-1}[x] + 
{\tt Cost}_{\tt Op}^t[{x^\ast}^{t}_i,a_i^t] + {\tt Cost}_{\tt Sw}^t[x,{x^\ast}^{t}_i]
\end{align}

\item {\bf Online Algorithm:}
To make online decisions for $({\sf P}_i)$, one can only rely on the present or past knowledge at timeslot $t$, without the future knowledge of $a_i^{t'}$, ${\tt p}^{t'}_{+}(j)$ and ${\tt p}^{t'}_{\mbox{-}}(j)$ for any $j$ and $t' > t$. Deciding an online decision is a hard problem, because of the presence of switching costs. Switching to a state with low current state cost may not offset the switching cost, when the state cost increases in the future.
An online algorithm for $({\sf P}_i)$ is known as {\em Work Function algorithm} (WFA) \cite{BEY05online}. The basic idea of WFA is to find a selection that minimizes the discrepancy with the offline optimal cost at the current timeslot $t$, subject to the constraint that the selection does not switch from the previous timeslot in the offline optimal cost. Formally, let $\hat{x}_{i}^t$ be the selection produced by WFA at timeslot $t$. $\hat{x}_{i}^t$ is decided by the following equation:
\begin{equation} \label{eqn:wfa}
\hat{x}_{i}^t = \arg\min_{x \in {\mathcal E}}\Big( {\tt Opt}_i^{t}[x] + {\tt Cost}_{\tt Sw}^t[{\hat{x}_i^{t-1}},x] \Big)
%({\tt c}_{x} +  {\tt d}_{\hat{x}_{i}^{t-1}} )\cdot {\mathds 1}_{\hat{x}_{i}^{t-1} \ne x} \Big)
\end{equation}
subject to
\begin{equation} \label{eqn:wfa-constraint}
{\tt Opt}^{t}_i[x] = {\tt Opt}^{t-1}_i[x] + {\tt Cost}_{\tt Op}^t[x,a_i^t]
%{\tt p}^t_{\hat{x}_i^t}[a_i^t] 
\end{equation}
where $\hat{x}_{i}^{t-1}$ is the selection of the previous timeslot by WFA.
Constraint~\raf{eqn:wfa-constraint} means that only the selections without switching at the previous timeslot in the offline optimal cost are the feasible candidates. Note that there always exists $x \in  {\mathcal E}$ that satisfies Constraint~\raf{eqn:wfa-constraint}, namely, at least one $x$ that does not switch from the previous timeslot in the offline optimal solution. Otherwise, ${\tt Opt}^{t}_i[x] $ is not optimal. We remark that the computation of ${\tt Opt}^{t}_i[x]$ does not need any knowledge of future timeslot $t' > t$. Hence, WFA is an online algorithm.

\end{enumerate}

Following the standard terminology of competitive analysis \cite{BEY05online}, we define the {\em competitive ratio} of an online algorithm by the worst-case ratio between the online solution and offline optimal solution: $\max\frac{{\tt Cost}_i[\hat{x}_{i}]}{{\tt Opt}^{T}_i[{x^\ast}^T_i] }$ over all possible inputs. There is a general upper bound of the competitive ratio of WFA.

\begin{customthm}{1}[\cite{BEY05online}]
If there are $n$ states in a MTS problem, then the competitive ratio of WFA is upper bounded by $2n-1$. Namely, in our problem, each state is an energy plan, and $n = |{\mathcal E}|$. Hence,
\[
{\tt Cost}_i[\hat{x}_{i}] \le (2 |{\mathcal E}| -1) \cdot {\tt Opt}^{T}_i[{x^\ast}^T_i] 
\]	
\end{customthm}

If there are only two energy plans, then we obtain ${\tt Cost}_i[\hat{x}_{i}] \le 3 \cdot {\tt Opt}^{T}_i[{x^\ast}^T_i]$. In fact, we can show a constant upper bound on the competitive ratio of WFA in a special case.

\begin{customthm}{2}
If the connection fees and disconnection fees of all energy plans are identical, i.e., ${\tt c}_{j} = {\tt c}$ and ${\tt d}_{j} = {\tt d}$ for all $j \in {\mathcal E}$, then 
\[
{\tt Cost}_i[\hat{x}_{i}] \le 3 \cdot {\tt Opt}^{T}_i[{x^\ast}^T_i] 
\]	
\end{customthm}

\noindent
Remarkably, even though the (dis)connection fees are not exactly identical, it is observed that the empirical competitive ratio is still well below the upper bound 3 in our evaluation.

\subsection{Group-based Online Energy Plan Selection} 

In this section, we incorporate group purchasing into our energy selection problem. We consider a group-based energy plan denoted by ${\mathfrak g}$, in addition to other individual energy plans ${\mathcal E} \backslash \{ {\mathfrak g}\}$. Group-based energy plan ${\mathfrak g}$ usually has more favorable tariffs. But there are other characteristics of a group-based energy plan in practice, such as the requirement of a minimum number of sign-up users:
\begin{enumerate}

\item {\bf\em Joining}: To join energy plan ${\mathfrak g}$, there requires a minimum number of ${\tt N}({\mathfrak g})$ sign-up users. One may interpret ${\tt N}({\mathfrak g})$ as the minimum aggregate bargaining power. When we set ${\tt N}({\mathfrak g}) = 1$, then any user can join ${\mathfrak g}$, without group purchasing. 
%Let the number of users currently using ${\mathfrak g}$ without termination at timeslot $t$ be $n^t({\mathfrak g})$. A new user can join ${\mathfrak g}$, but there requires at least a number of ${\tt N}({\mathfrak g}) - n^t({\mathfrak g})$ new users to sign up at $t$, in addition to the number of $n^t({\mathfrak g})$ current users.

\item {\bf\em Termination}:  We consider a simple setting of termination. Any user can unilaterally terminate plan ${\mathfrak g}$ before the contract period by paying the disconnection fee ${\tt d}({\mathfrak g})$, which may be higher than any individual energy plan.

%\item {\em Maximum Discounted Period}: Let ${\tt D}({\mathfrak g})$ be the maximum period that ${\mathfrak g}$ will last. Beyond the period ${\tt D}({\mathfrak g})$, no one can join the group plan any more.

\end{enumerate}

We next extend the online algorithm WFA to the setting of group purchasing for energy plans. Suppose ${\mathfrak g} \in {\mathcal E}$ is the only group-based plan (i.e., ${\sf N}({\mathfrak g}) > 1$), while the rest are individual energy plans (i.e., ${\sf N}(j) = 1$, for all $j \in {\mathcal E} \backslash \{ {\mathfrak g} \}$). Given a subset of users $X \subseteq {\mathcal N}$ (who have not joined group-based plan ${\mathfrak g}$), we will decide whether $X$ should join group energy plan ${\mathfrak g}$ or not. If so, how to meet the requirement of minimum number of users when sign-up.

Let $(\hat{x}_{i}^{t-1})_{i \in X}$ be the users' selections of the previous timeslot $t-1$. Naturally, each user $i \in X$ applies WFA to the two following sub-problems:
\begin{enumerate}

\item Considering only group-based plan ${\mathfrak g}$:  
If ${\tt Opt}^{t}_i[{\mathfrak g}] \ne {\tt Opt}^{t-1}_i[{\mathfrak g}] + {\tt Cost}_{\tt Op}^t[{g},a_i^t]$ for any user $i \in X$ (namely, Constraint~\raf{eqn:wfa-constraint} is violated), then ${\mathfrak g}$ will not be considered. Otherwise, let 
\[
C^t_i({\mathfrak g}) \triangleq  {\tt Opt}_i^{t}[{\mathfrak g}] +  {\tt Cost}_{\tt Sw}^t[{\hat{x}_i^{t-1}},g] 
\]

\item Considering the rest of energy plans ${\mathcal E} \backslash \{ {\mathfrak g} \}$: Find $y_i \in {\mathcal E} \backslash \{ {\mathfrak g} \}$, such that
\[
y_i = \arg\min_{y \in {\mathcal E} \backslash \{ {\mathfrak g} \}}\Big( {\tt Opt}_i^{t}[y] + {\tt Cost}_{\tt Sw}^t[{\hat{x}_i^{t-1}},y]   \Big)
\]
subject to \	 ${\tt Opt}^{t}_i[y] = {\tt Opt}^{t-1}_i[y] + {\tt Cost}_{\tt Op}^t[{y},a_i^t]$. Let 
$$C^t_i({\mathcal E} \backslash \{ {\mathfrak g} \}) \triangleq {\tt Opt}_i^{t}[y_i] +  {\tt Cost}_{\tt Sw}^t[{\hat{x}_i^{t-1}},y_i]$$
Let $C^t_i({\mathcal E} \backslash \{ {\mathfrak g} \}) = \infty$, if no $y$ satisfies Constraint~\raf{eqn:wfa-constraint}.

\end{enumerate}

Let us first consider the setting where each user decides separately whether to join group-based plan ${\mathfrak g}$. Let $m_X^t({\mathfrak g})$ be the number of users in $X$ who strictly prefer ${\mathfrak g}$, namely,
\[
m_X^t({\mathfrak g}) \triangleq \Big|\big\{ i \in X \mid  C^t_i({\mathfrak g}) \le C^t_i({\mathcal E} \backslash \{ {\mathfrak g} \})\big\} \Big|
\]

If $m_X^t({\mathfrak g}) \ge {\sf N}({\mathfrak g})$, then these $m_X^t({\mathfrak g})$ users will naturally join energy plan ${\mathfrak g}$. However, if $m_X^t({\mathfrak g}) < {\sf N}({\mathfrak g})$, then these users are unable to join energy plan ${\mathfrak g}$, without the required minimum number of sign-up users. In the next section, we will incorporate {\em mutual compensations} to incentivize users to join energy plan ${\mathfrak g}$ by compensating their switching costs of terminating their current energy plans, when $m_X^t({\mathfrak g}) < {\sf N}({\mathfrak g})$.

\subsection{\mbox{Mutual Compensations for Group Purchasing}}

In this section, we consider the possibility that some users will pay mutual compensations to other users for partially subsidizing their switching costs in order to join group-based plan ${\mathfrak g}$. 

It is not always true that $C^t_i({\mathfrak g}) \le C^t_i({\mathcal E} \backslash \{ {\mathfrak g} \})$ for all $i \in X$, as some users may have $C^t_i({\mathcal E} \backslash \{ {\mathfrak g} \}) < C^t_i({\mathfrak g})$. However, if $\sum_{i \in X} C^t_i({\mathfrak g}) \le \sum_{i \in X} C^t_i({\mathcal E} \backslash \{ {\mathfrak g} \})$ (i.e., the sum, instead of individual users), then some users may compensate others in order to join energy plan ${\mathfrak g}$. Let $\theta_i$ be the {\em compensated switching cost} for user $i$. Hence, rather than considering $C^t_i({\mathfrak g})$ in WFA, each user $i$ considers the following compensated cost:
\begin{equation}
{\tt Opt}_i^{t}[{\mathfrak g}] + \theta_i 
\end{equation}
Note that some users may pay more (i.e., $\theta_i  > {\tt Cost}_{\tt Sw}^t[{\hat{x}_i^{t-1}},g]$), while other users may pay less (i.e., $\theta_i  < {\tt Cost}_{\tt Sw}^t[{\hat{x}_i^{t-1}},g]$).

To find proper values for $(\theta_i )_{i \in X}$, we consider several properties:
\begin{itemize}

\item {\bf\em Individual Rationality}: With compensated switching cost, each user can apply WFA individually to decide whether to join group-based plan ${\mathfrak g}$ in an online manner. All users prefer group-based plan ${\mathfrak g}$, if the following condition of {\em individual rationality} is satisfied for all $i \in X$:
\begin{equation}
{\tt Opt}_i^{t}[{\mathfrak g}] + \theta_i < C^t_i({\mathcal E} \backslash \{ {\mathfrak g} \})
\end{equation}

\item {\bf\em Budget Balance}: We consider no external financial support, namely, the total compensated switching cost should equal the default switching cost without mutual compensations. Hence, the following condition of {\em budget balance} is required:
\begin{equation}
\sum_{i \in X}  \theta_i  =  \sum_{i \in X}  {\tt Cost}_{\tt Sw}^t[{\hat{x}_i^{t-1}},g]
\end{equation}
Equivalently, $\sum_{i \in X} \Big( {\tt Opt}_i^{t}[{\mathfrak g}] + \theta_i \Big) = \sum_{i \in X} C^t_i({\mathfrak g})$.

\item  {\bf\em Group Feasibility}: From the perspective of a social planner, we also consider the following social online decision problem with aggregate cost of all users, as defined as follows:
\begin{equation} \label{eqn:mincost}
\begin{array}{lc}
 ({\sf P}_{\rm soc}) \ \ \ & \displaystyle \min_{x \in {\mathcal E}^{T|X|}} \sum_{i \in X} {\tt Cost}_i[x_{i}]
 \end{array}
\end{equation}
where $x = (x_i)_{i \in X}$ is the collective selections of all users, and ${\mathcal E}^{T|X|}$ is the feasible space. Unlike the individual problem $({\sf P}_i)$, the social problem $({\sf P}_{\rm soc})$ minimizes the total long-term total energy cost of all users in $X$.
When we apply WFA to problem $({\sf P}_{\rm soc})$, the online decision to switch to group-based plan ${\mathfrak g}$ is feasible, if the following condition of {\em group feasibility} is satisfied:
\begin{equation}
\sum_{i \in X} C^t_i({\mathfrak g}) < \sum_{i \in X} C^t_i({\mathcal E} \backslash \{ {\mathfrak g} \})
\end{equation}

\end{itemize}

In practice, individual rationality and budget balance are important properties of a feasible scheme of mutual compensations, whereas group feasibility can be easily checked before deciding mutual compensations.

\begin{customthm}{3}\label{thm:payments}

We consider two schemes of mutual compensations (based on the ideas
in \cite{p2p19,CE20sharing,CE17sharing}):
\begin{enumerate}

\item{\bf Egalitarian Cost-sharing:}
\begin{align} 
\theta_i  =\  & C^t_i({\mathcal E} \backslash \{ {\mathfrak g} \})  +  \frac{ \sum_{i' \in X} \big( C^t_{i'}({\mathfrak g}) - C^t_{i'}({\mathcal E} \backslash \{ {\mathfrak g} \})  \big)}{|X|} - {\tt Opt}_i^{t}[{\mathfrak g}] 
\end{align}

\item{\bf Proportional Cost-sharing:}
\begin{align} 
\theta_i  =\  & C^t_i({\mathcal E} \backslash \{ {\mathfrak g} \}) \frac{\sum_{i' \in X} C^t_{i'}({\mathfrak g})}{\sum_{i' \in X} C^t_{i'}({\mathcal E} \backslash \{ {\mathfrak g} \}) } - {\tt Opt}_i^{t}[{\mathfrak g}] 
\end{align}

\end{enumerate}
Then $(\theta_i )_{i \in X}$ in both schemes satisfy individual rationality and budget balance, when group feasibility is satisfied. 
\end{customthm}

See Appendix~\ref{sec:append1} for the proof.
Theorem~\ref{thm:payments} provides two feasible schemes of mutual compensations that guarantee individual rationality and budget balance, given group feasibility. Hence, it suffices to consider the social problem $({\sf P}_{\rm soc})$ and apply WFA to $({\sf P}_{\rm soc})$ with respect to $X$.

\subsection{\mbox{Group Purchasing Decision-Making Mechanism}} \label{sec:decision-mech}

\begin{figure}[ht!] 
\includegraphics[width=\columnwidth]{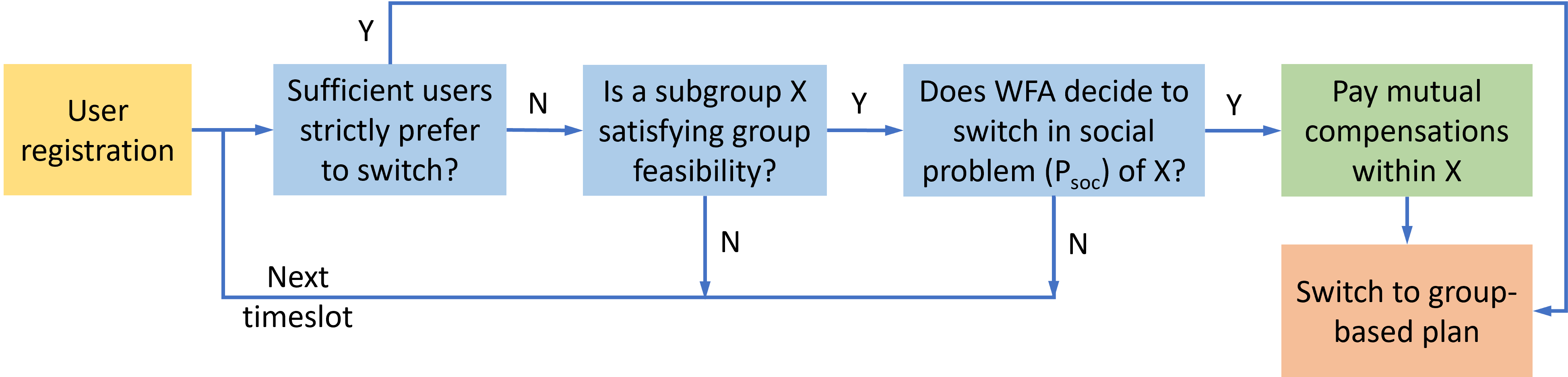} 
\caption{A flowchart of the group purchasing decision-making mechanism.}
\label{fig:flowchart}
\end{figure}

Based on the above results, this section presents a decision-making mechanism for group purchasing, incorporating mutual compensations. As illustrated in Figure~\ref{fig:flowchart}, we determine the group-purchasing decision as follows:

\begin{longfbox}[border-break-style=none,border-color=\#bbbbbb,background-color=\#eeeeee,breakable=true,width=\linewidth]

{\bf Group Purchasing Decision-Making Mechanism}:
\begin{enumerate}

\item The users who are interested in joining group-based plan ${\mathfrak g}$ first register as ${\mathcal N}$. They agree on whether egalitarian cost-sharing or proportional cost-sharing will be used.

\item At each timeslot $t$, check if there are sufficient users who strictly prefer to switch to group-based plan ${\mathfrak g}$ without mutual compensations. Namely, if $m_X^t({\mathfrak g}) \ge {\sf N}({\mathfrak g})$, then these $m_X^t({\mathfrak g})$ users will join group-based plan ${\mathfrak g}$.

\item Otherwise, if $m_X^t({\mathfrak g}) < {\sf N}({\mathfrak g})$
\begin{enumerate}

\item Find a subset $X \subseteq {\mathcal N}$, such that $|X| \ge {\sf N}({\mathfrak g})$ and group feasibility is satisfied within $X$, namely, $\sum_{i \in X} C^t_i({\mathfrak g}) < \sum_{i \in X} C^t_i({\mathcal E} \backslash \{ {\mathfrak g} \})$. Apply WFA on $({\sf P}_{\rm soc})$ with respect to $X$.

\item If WFA decides to switch to group-based plan ${\mathfrak g}$ in $({\sf P}_{\rm soc})$ with $X$, then compute $(\theta_i )_{i \in X}$ of the respective mutual compensation scheme. Based on Theorem~\ref{thm:payments}, individual rationality and budget balance are guaranteed.

\item After paying the respective mutual compensations to each other, the users in $X$ will join group-based plan ${\mathfrak g}$. 

\end{enumerate}
\end{enumerate}

\end{longfbox}

\medskip

The above decision-making mechanism can be conducted in a centralized manner or via a third-party, when every user discloses their private energy data. In the rest of the paper, we seek to perform the decision-making mechanism in a decentralized privacy-preserving manner, without disclosing private energy data.

\iffalse
\begin{customthm}{4}
If we apply WFA to problem $({\sf P}_{\rm soc})$ to produce online decision $\hat{x}$. Then the competitive ratio is upper bounded by
\[
{\tt Cost}[\hat{x}] \le (2 |{\mathcal E}| -1) \cdot {\tt Opt}^T[{x^\ast}^T] 
\]	
where ${x^\ast}^T$ is an offline optimal solution of $({\sf P}_{\rm soc})$.
\end{customthm}
\fi

%% file: model2.tex
\section{Security \& Threat Model}  \label{sec:model-security}

In this section, we formulate privacy protection in the group purchasing decision-making mechanism. Note that we do not consider privacy protection in the communication. We assume that secure, reliable and authenticated communications can be established among the parties, with no man-in-the-middle attack. Secure communications can be attained by suitable end-to-end security.

\subsection{Security and Privacy Requirements} \label{sec:req}

In this paper, we provide a privacy-preserving solution to enable the users to jointly conduct the decision-making mechanism in Section~\ref{sec:decision-mech} to determine the group purchasing decisions and mutual compensations, given their demands and current energy plans. Our privacy-preserving  solution achieves the following security and privacy requirements in the decision-making mechanism:
\begin{enumerate}

\item {\bf Private Demands:}  No users will leak their past demands $(a_i^t)$ to other users. However, users will be able to verify their past demands from an authorized source, without revealing their demands to each other.

\item {\bf Private Selections:} No users will leak their currently selected energy plans $(x_i^t)$ to other users, if they do not need to share the same group-based energy plan.

\item {\bf Private Compensations:}  No users will leak their mutual compensations received from other users $(\theta_i^t)$, because it may reveal their demands or energy plans.

\end{enumerate}

Nonetheless, we assume that all energy plan information ${\tt E}(j) = \langle {\tt p}^t_+(j), {\tt p}^t_{\mbox{-}}(j), {\tt c}(j), {\tt d}(j),  {\tt T}(j)\rangle$ is publicly known for all $j \in {\mathcal E}$.

\subsection{Threat Model}

Note that dishonest user may take advantage of privacy protection to perform the following attacks in the decision-making process:

\begin{enumerate}

\item {\bf\em Misrepresentation:} Dishonest users may try to misrepresent their demands or current energy plans in the decision-making process.

\item {\bf\em Mis-computation:} They may output incorrect or inconsistent values in any joint computation process (e.g., validation or generation of proofs).

\item {\bf\em Incorrect Compensations:}  They may try to cheat by paying less compensation to others, or claim more compensation than what they ought to. 

\end{enumerate}
Dishonest users may also collude with each other to coordinate their actions. To facilitate our solution, we make some assumptions about the dishonest users:

\begin{enumerate}

\item {\em Maximum number:} Up to $(N-2)$ dishonest users, who may be adaptive adversaries and can deliberately deviate from the protocols for prying into others' privacy or sabotage the protocols. Their disruptive actions can happen at any time during the protocols rather than before the protocols (as for static adversaries). In case of any dishonest actions being detected, our system will abort and notify all the users. 

\item {\em Identifiability:} Our solution ensures that any dishonest actions will be detected. However, our system is not required to identify individual dishonest users, as it is fundamentally impossible \cite{BGW99} to identify a dishonest user in multi-party computation with a majority of dishonest users. There are secure multi-party computation protocols \cite{cramer2001multiparty} that can identify a dishonest user, but requiring a majority of honest users and considerable computational overhead. On the other hand, we may impose further measures to mitigate dishonesty. For example, requiring proper user authentication to prevent shilling. Or, we can require each user to pay a deposit in advance, which will be forfeited if any dishonesty is detected.

\end{enumerate}

\section{System Model} \label{sec:model-blockchain}

In this section, we present the system models of blockchain and energy bill data that underlie  our privacy-preserving solution.

\subsection{Blockchain Model}

We consider an account-based blockchain model like Ethereum (which is a general-purpose blockchain platform \cite{ethereum}), whereas Bitcoin operates with a different transaction-output model for cryptocurrency transactions only. Smart contracts are programmable code on a blockchain platform that provide customized computation tasks along with each transaction (e.g., transaction validation, data processing). Each user has an account associated with the blockchain, which allows them to pay mutual compensations to each other. The encrypted energy bill data and group purchasing decisions are also recorded on the blockchain for later validation.

The blockchain consists of several components:
\begin{enumerate}

\item {\bf Ledger}: An append-only ledger on a blockchain holds the records of all accounts and transactions. Note that by default, there is no privacy protection to the ledger, such that the account details and transaction histories are visible to the public. On Ethereum, one can create tokens (ERC20 \cite{erc20}) on the ledger to represent certain digital assets. Our mutual compensation payment system is implemented by ERC20 tokens, which allows us to incorporate privacy protection. To make payment among each other, users are required to purchase tokens that will be subsequently transferred to each other and redeemed.  

\item {\bf Accounts}: An account is identified by a public key $K^{\tt p}$ and an address ${\tt ad}$, which is the hash of the public key: ${\tt ad} = {\mathcal H}(K^{\tt p})$, where ${\mathcal H}(\cdot)$ is a cryptographic hash function. The user manages the account by a private key $K^{\tt s}$. Each account holds a balance of tokens, denoted by ${\tt Bal}({\tt ad})$, which by default is a publicly visible plaintext. Each user $i$ has an account associated with a tuple $({\tt ad}_i, K^{\tt p}_i, K^{\tt s}_i, {\tt Bal}({\tt ad}_i))$. 

\item {\bf Transactions}: To initiate a transaction of tokens from ${\tt ad}_i$ to ${\tt ad}_{i'}$ with transaction value ${\tt val}$, a user submits a transaction request to the blockchain: ${\tt tx} = ({\tt ad}_i, {\tt ad}_{i'}, {\tt val})$, along with a signature ${\tt sign}_{K^{\tt s}_i}({\tt tx})$ using the private key $K^{\tt s}_i$ associated with ${\tt ad}_i$. The transaction request will be executed\footnote{A blockchain transaction has more security elements, like nonce to prevent replay attack, account-locking against front-running attack, etc. But our model can easily incorporate these elements in practice.} if ${\tt Bal}({{\tt ad}_i}) \ge {\tt val}$. 

\item {\bf Data}: A small amount of data can also be recorded by smart contracts on the blockchain ledger, which is visible to the public. We will introduce a privacy-preserving compact approach of data representation by cryptographic commitments and Merkel trees in the next section.

\end{enumerate}

%Note that there may be a negative flow of payment in egalitarian mutual compensation, such that ${\tt val}_i < 0$. Hence, we need to ensure the corresponding transaction on a blockchain still functions correctly. 

\subsection{Energy Bill Data}

\subsubsection{Cryptographic Commitments}
\

A (cryptographic) commitment allows a user to hide a secret (e.g., to hide the balances and transactions on a blockchain), as known as Pedersen commitment. Denote a finite field by ${\mathbb Z}_p$. To commit secret value $x\in{\mathbb Z}_p$, a user first picks a random number ${\tt r} \in {\mathbb Z}_p$ to mask the commitment, and then computes the commitment by: 
\begin{equation}
\label{eqn:pedersen}
{\tt Cm}(x, {\tt r}) = g^x \cdot h^{\tt r} \  (\mbox{mod\ } p)
\end{equation}
where $g, h$ are generator of a multiplicative group ${\mathbb Z}_p^*$, and $p$ is a large prime number. Pedersen commitment is perfectly hiding (i.e., no adversary can unlock the secret) and computationally binding (i.e., no adversary can associate with another secret in polynomial time). Note that Pedersen commitment satisfies homomorphic property: ${\tt Cm}(x_1+x_2, {\tt r}_1+{\tt r}_2) = {\tt Cm}(x_1, {\tt r}_1) \cdot {\tt Cm}(x_2, {\tt r}_2)$. Sometimes, we simply write ${\tt Cm}(x)$ without specifying random ${\tt r}$. 

\subsubsection{Merkel Trees}
\

\begin{figure}[!h]
\centering  
\includegraphics[width=0.5\linewidth]{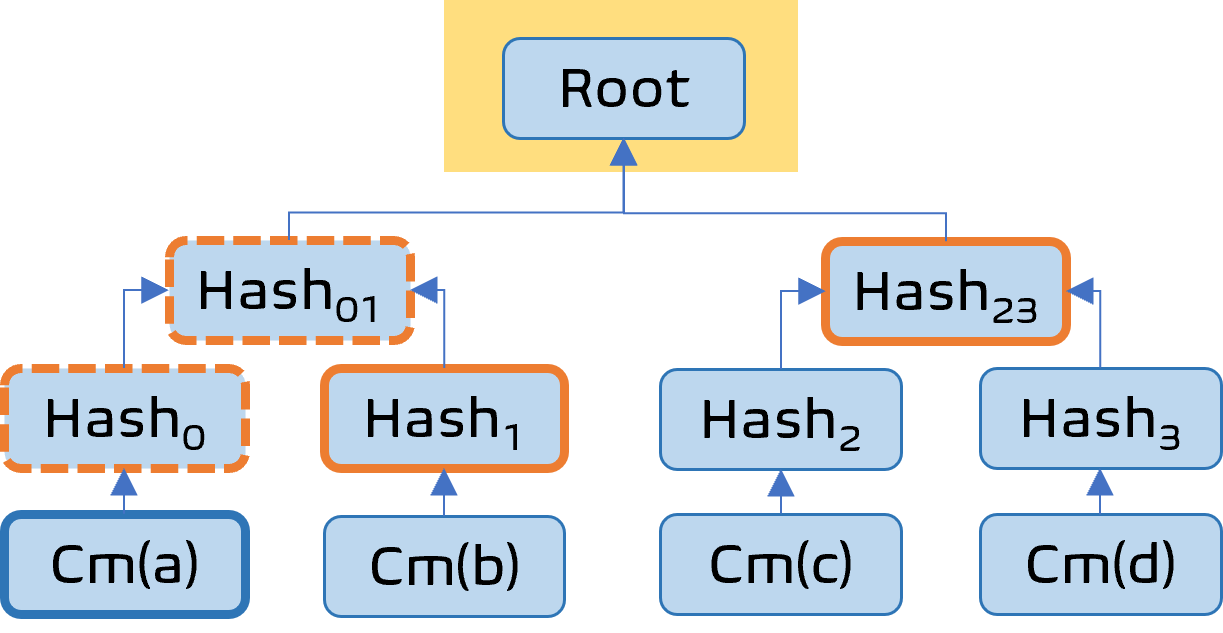} \vspace{-5pt}
\caption{An example of a given Merkle tree and a proof of membership for ${\tt Cm}(a)$ in the Merkle tree is $({\tt Hash_1}, {\tt Hash_{23}})$, such that ${\tt Root} = {\mathcal H}( {\mathcal H}( {\mathcal H}({\tt Cm}(a))\mid {\tt Hash_1}) \mid {\tt Hash_{23}})$} 
\label{fig:merkle} \vspace{-10pt}
\end{figure}

To reduce the ledger storage space, we use a Merkel tree for data representation and only a small-sized hash pointer (i.e., the root of the Merkel tree) will be recorded on the ledger. A Merkle tree is a binary tree of hash pointers, such that the data value of a non-leave node is ${\tt Hash_{XY}} \triangleq {\mathcal H}({\tt X} \mid {\tt Y})$, where ${\tt X}, {\tt Y}$ are the data values of its left and right children respectively (which can also be hash pointers). A hash pointer is a concise way of data representation. The root of Merkle tree, denoted by {\tt Root} represents the hash pointer of the whole tree. We associate each leaf in the Merkle tree by a commitment. Note that only a concise proof is needed to prove that a given data value of a leaf is a member in a given Merkle tree. For example, to prove the membership of a leaf, ${\tt Cm}(a)$, in the Merkle tree in Figure~\ref{fig:merkle} with ${\tt Root}$, one only needs to locally record ${\tt Hash_1}, {\tt Hash_{23}}$ as a proof, highlighted in bold orange nodes in Figure~\ref{fig:merkle}, with which one can respectively reconstruct the hashes $({\tt Hash_0}, {\tt Hash_{01}})$, up to ${\tt Root}$. Merkle tree eliminates the need of storing the entire dataset, but only the root. In this case, the prover only needs to record an $O(\log n)$-sized proof for each data value.

\subsubsection{Encrypted Energy Bill Receipts}
\

We next use commitments and Merkle trees to record the encrypted energy bill receipts. Recall that each user $i$ has energy demands $a_i^t$ at timeslot $t$. Let an indicator variable be $\beta^t_i \triangleq {\mathds 1}\{a_i^t\ge 0\}$, and the corresponding operational energy cost at timeslot $t$ be $\kappa_{i}^t \triangleq {\tt Cost}_{\tt Op}^t[x^{t}_i,a_i^t]$.
Let the potential connection and disconnection fees of each user $i$'s current energy plan at timeslot $t$ be
$$
\mu_{i}^t \triangleq  {\tt c}({x_i^t}), \qquad
\nu_{i}^t  \triangleq  {\tt d}({x_i^{t-1}}) \cdot {\mathds 1}\{{\tt T}_i^t \le {\tt T}({x_i^{t-1}})\} 
$$

After the user paying the energy bill, the utility operator generates an encrypted receipt as follows: For a fixed epoch $[t_0, t_1]$, the utility operator records the following tuple for each timeslot $t \in [t_0, t_1]$ of the user:
$$
\Big( {\tt Cm}(a^t_i), {\tt Cm}(\beta^t_i), {\tt Cm}(\kappa^t_i), {\tt Cm}(\mu_{i}^t), {\tt Cm}(\nu_{i}^t) \Big)_{t = t_0}^{t_1}
$$
These tuples will be represented by leaves in a Merkle tree, and its root with a verifiable signature and a timestamp will be recorded on the ledger. In the next sequent, we will utilize multi-party computation for validating if the shared energy bills data matches the encrypted receipts on the ledger.

%% file: spdz.tex
\section{Multi-party Computation Protocol}  \label{sec:spdz}

We presented a decision mechanism for group purchasing in Section~\ref{sec:decision-mech}. In this section, we describe the way to execute the mechanism in a decentralized privacy-preserving manner by multi-party computation. This section presents a simplified multi-party computation protocol called SPDZ \cite{cramer2015secure, dklpss13spdz}, which allows multiple parties to jointly compute a certain function with concealed input. SPDZ can safeguard against a dishonest majority (i.e., all but one party may be dishonest), and does not require a trusted setup.

The full details of SPDZ protocol can be found in Appendix.~\ref{sec:append3}. Here, we only sketch some high-level ideas of SPDZ, and show how to use SPDZ to validate secretly shared private input by cryptographic commitments on the blockchain ledger.

\subsection{Basic Operations of SPDZ} 

In a nutshell, SPDZ relies on secret-sharing, whereby private data will be distributed to multiple parties, such that each party only knows a share of the data, without complete knowledge of other shares. Thus, the computation of individual shares of data will not reveal the original data, unless all shares are revealed for output or validation. Several distributed computation operations can be performed locally, while preserving the secret-sharing property. 

Suppose a private number $x$ is distributed to $n$ parties, such that each party $i$ knows a share $x_i$ only and $x = \sum_{i=1}^n x_i$, but not knowing other shares $x_j$ for $j \ne i$. Note that no one is unable to construct $x$, without knowing all the shares. In the following, we write $\langle x \rangle$ as a {\em secretly shared} number, meaning that there is a vector $(x_1, ..., x_n)$, such that each party $i$ knows only $x_i$. Given secretly shared $\langle x \rangle$ and $\langle y \rangle$, and a public known constant $c$, we can compute the following operations in a privacy-preserving manner by local computation at each party, and then the outcome can be assembled from the individual shares: 
\begin{enumerate}

\item[{\tt A1})] $\langle x \rangle + \langle y \rangle$ can be computed locally by $(x_1 + y_1, ..., x_n + y_n)$.

\item[{\tt A2})] $c \cdot \langle x \rangle$ can be computed locally by $(c \cdot x_1, ..., c \cdot x_n)$.

\item[{\tt A3})] $c + \langle x \rangle$ can be computed locally by $(c + x_1, x_2, ..., x_n)$.

\end{enumerate}
To reveal $\langle x \rangle$, each party $i$ simply broadcasts $x_i$ to other parties. Then each party can reconstruct $x = \sum_{i=1}^n x_i$. Note that multiplications can also be computed in a privacy-preserving manner, albeit with a more complex setup (see Appendix.~\ref{sec:append3}). With additions and multiplications, one can compute a large class of functions (including comparison and branching conditions).

Note that some parties may be dishonest, who may not perform the required local computations correctly. We need to safeguard against dishonest parties by validation through message authentication codes (MACs). Every secretly shared number is encoded by a MAC as $\gamma(x)$, which is also secretly shared as $\langle \gamma(x) \rangle$. The basic idea is that if a dishonest party wants to modify his share $x_i$, then he also needs to modify $\gamma(x)_i$ consistently. This allows dishonesty to be detectable by checking the corresponding MAC in the final output. In the following, we write $\llangle  x \rrangle$ meaning that both $\langle  x \rangle$ and the respective MAC $\langle \gamma(x) \rangle$ are secretly shared among the users.

We next sketch a high-level description of SPDZ protocol:
\begin{enumerate}

\item {\em Pre-processing Phase}: In this phase, a collection of shared random numbers will be constructed that can be used to mask the private input numbers. For each private input number of party $i$, there needs a shared random number $\llangle r^i \rrangle$, where $r^i$ is revealed to party $i$ only, but not to other parties.

\item {\em Online Phase}: To secretly shares a private input number $x^i$ using $\llangle r^i \rrangle$, without revealing $x^i$, it proceeds as follows:
\begin{enumerate}

\item[1)] Party $i$ computes and reveals $z^i = x^i - r^i$ to all parties.

\item[2)] Every party sets $\llangle x^i \rrangle \leftarrow z^i + \llangle r^i \rrangle$ (see {\tt A3}).

\end{enumerate}

Next, any computation functions in terms of additions or multiplications can be computed by proper local computations (e.g., {\tt A1}-{\tt A3}). The MACs are updated accordingly to preserve the consistency. See Appendix.~\ref{sec:append3} for details.

\item {\em Output and Validation Phase}: All MACs will be revealed for validation. If there is any inconsistency in MACs, then abort.

\end{enumerate}

\subsection{\mbox{Decentralized Validation of Energy Bill Data}} \label{sec:validatebill}

With SPDZ, one can perform the decision-making mechanism of Section~\ref{sec:decision-mech} in a decentralized privacy-preserving manner. We will provide the detailed procedures in SPDZ in the next section. In the following, we particularly explain the joint validation of encrypted private input data on the blockchain ledger via SPDZ.

On one hand, each user $i$ secretly shares her energy bill data $\big(\llangle a^t_i\rrangle,$ $\llangle\beta^t_i \rrangle, \llangle\kappa^t_i \rrangle, \llangle\mu_{i}^t \rrangle, \llangle \nu_{i}^t \rrangle\big)$ as the input to the decision-making mechanism. On the other hand, user $i$ proves that the corresponding $\big({\tt Cm}(a^t_i), {\tt Cm}(\beta^t_i), {\tt Cm}(\kappa^t_i), {\tt Cm}(\mu_{i}^t),$ ${\tt Cm}(\nu_{i}^t)\big)$ are recorded by a root of a Merkle tree on the ledger with appropriate signature and timestamp from the utility operator. The proof of a membership in a Merkle tree is attained by providing a valid path in the Merkle tree that matches the root. 

However, there may be inconsistency between these inputs of a dishonest user, namely, the secretly shared values may not match the commitments. Hence, to ensure the consistency, the users generate ZKPs to prove the knowledge in the commitments from the secretly shared values. For example, user $i$ generates a ZKP for the proof of knowledge of $a^t_i$ in ${\tt Cm}(a^t_i)$ from $\llangle a^t_i\rrangle$, without disclosing $a^t_i$. Note that there is a well-known technique (called $\Sigma$-protocol; see Appendix~\ref{sec:append2}) to generate a ZKP from a private value $x$ for a given ${\tt Cm}(x, {\tt r})$. We briefly review the its construction as follows:
\begin{enumerate}

\item
 The prover generates a pair of random numbers $(x', {\tt r}')$ and announces its commitment ${\tt Cm}(x', {\tt r}')$ to the verifier.
		
\item
The verifier generates a random challenge $\psi$ and announces $\psi$ to the prover.

\item
The prover computes $z_x \leftarrow x' + \psi \cdot x$ and $z_{\tt r} \leftarrow {\tt r}' + \psi \cdot {\tt r}$, and announces $(z_x, z_{\tt r})$ to the verifier.

\item
The verifier checks the condition:
If $g^{z_x} \cdot h^{z_{\tt r}} =  {\tt Cm}(x', {\tt r}') \cdot {\tt Cm}(x, {\tt r})^\psi$ (which is based on the homomorphic property of Pedersen commitments), then it passes the verification.

\end{enumerate}
 
Next, we replace the above construction of ZKP by a distributed version via SPDZ, denoted by $\Pi_{\tt dzkpCm}$ (see Algorithm~\ref{alg:dzkpCm}). Since $\Pi_{\tt dzkpCm}$ is carried out jointly by all users without disclosing the secretly shared input, passing the verification will justify the consistency between the secretly shared values and the commitments. In $\Pi_{\tt dzkpCm}$, the random challenge $\psi$ is generated by random strings from all users. Also, the MACs in SPDZ will ensure that the distributed computations of $\Pi_{\tt dzkpCm}$ are performed correctly, despite the presence of dishonest users.

\begin{algorithm}[tbh!] 
	\caption{$\Pi_{\tt dzkpCm}$: {\em Prove the knowledge of $x$ in a given commitment ${\tt Cm}(x)$ from secretly shared $\llangle x \rrangle$ via SPDZ}} \label{alg:dzkpCm}	
{\scriptsize	
	\begin{algorithmic}[1]		
		\Require ${\tt Cm}(x, {\tt r})$ (known to all users), $\llangle x \rrangle$ (already secretly shared among users); $(x,{\tt r})$ (known to user $i$ only)
		\Ensure {\sf Pass} or {\sf Fail}

		\State 
		User $i$ announces ${\tt Cm}(x, {\tt r})$ and secretly shares $\llangle {\tt r}\rrangle$ with all users

		\State 
		User $i$ generates a pair of random numbers $(x', {\tt r}')$ and announces ${\tt Cm}(x', {\tt r}')$ to all users 
		
		\State
		User $i$ secretly shares $\llangle x'\rrangle$ and $\llangle {\tt r}'\rrangle$ with all users		
		
		\State 
		All users conduct a coin-tossing protocol to obtain a random challenge $\psi$ as follows:
		
		\begin{enumerate}[leftmargin=15pt]
		
		\item 
		Each user $j$ announces a commitment ${\tt Cm}({\tt r}''_j)$ of a random string ${\tt r}''_j$ 
		\item 
		After receiving the commitments $\{{\tt Cm}({\tt r}''_j)\}$ from all users, each user reveals ${\tt r}''_j$ to all users and all users check if it matches ${\tt Cm}({\tt r}''_j)$
		\item 			
		\mbox{Set $\psi\leftarrow {\mathcal H}( {\tt r}''_1 | ... | {\tt r}''_{|X|})$ \Comment{\em Generate random challenge by hashing concatenated random strings}}
		\end{enumerate}
		
		\State  Compute the following via SPDZ among all users:
		\begin{align}
		\llangle z_{x} \rrangle \leftarrow \llangle x' \rrangle + \psi \cdot \llangle x\rrangle \notag\\
		\llangle z_{\tt r} \rrangle \leftarrow \llangle {\tt r}' \rrangle + \psi \cdot \llangle {\tt r} \rrangle \notag
		\end{align}

		\State Reveal $z_{x}$, $z_{\tt r}$ and their MACs to all users

		\Statex \LeftComment{{\em All users check the following}}
		
		\If{$g^{z_{x}} \cdot h^{z_{{\tt r}}} =  {\tt Cm}(x', {\tt r}')  \cdot {\tt Cm}\big(x, {\tt r}\big)^\psi$ and checking all revealed MACs passed}
		
		\State \Return {\sf Pass}

		\Else

		\State \Return {\sf Fail} and abort 
		
		\EndIf

		\Statex \LeftComment{\mbox{\em The distributed ZKP of commitment is ${\sf zkpCm}[x, {\tt Cm}(x)] = \big\{ {\tt Cm}(x, {\tt r}); {\tt Cm}(x', {\tt r}'), ({\tt r}''_i)_{i \in X}, z_x, z_{\tt r} \big\}$}}
		
	\end{algorithmic}	
}	
\end{algorithm}

%% file: algo.tex
\section{Privacy-Preserving Mechanism and Protocols} \label{sec:algo}

Based on the results of the previous sections, this section presents the full decision-making mechanism for group purchasing with privacy-preserving protocols based on SPDZ and ZKPs. For the brevity of presentation, we use the following shorthand notations in Table~\ref{tbl:symbs} to represent various energy costs and fees.

\begin{table}[tbh!]
    \centering
	{\scriptsize
    \begin{tabularx}{\linewidth}{@{ }r@{ }|@{ }c@{ }|@{ }l@{}}
	\hline  \hline   
	 Meaning & Notation & Representation \\
    \hline   
    Consumption indicator & $\beta^t_i$  & $={\mathds 1}\{a_i^t\ge 0\}$ \\ %\hline
    Operational energy cost of plan ${x_i^t}$ & $\kappa_{i}^t$ & $={\tt Cost}_{\tt Op}^t[x^{t}_i,a_i^t]$ \\ %\hline
   Connection fee of plan ${x_i^t}$ & $\mu_{i}^t$ & $={\tt c}({x_i^t})$ \\ %\hline
   Disconnection fee of plan ${x_i^t}$ & $\nu_{i}^t $ & $={\tt d}({x_i^{t-1}}) \cdot {\mathds 1}\{{\tt T}_i^t \le {\tt T}({x_i^{t-1}})\}$ \\   %\hline
   Operational energy cost of plan ${\mathfrak g}$ & $\kappa_{i,{\mathfrak g}}^t$ & $={\tt Cost}_{\tt Op}^t[{\mathfrak g},a_i^t]$ \\ %\hline
   Offline optimal cost of plan ${x_i^t}$ at $t$ & ${\tt Opt}_{i}^t$ & $={\tt Opt}_{i}^t[x_{i}^t]$ \\ %\hline
   Offline optimal cost of plan ${\mathfrak g}$ at $t$ & ${\tt Opt}_{i,{\mathfrak g}}^t$ & $={\tt Opt}_{i}^t[{\mathfrak g}]$ \\ 
    \hline  \hline    
    \end{tabularx}
	}
	\caption{\label{tbl:symbs}Table of shorthand notations.}  \vspace{-20pt}
\end{table}

To simplify the presentation, we only consider the switching decision for a group of users $X$ from their individual energy plans to a group-based plan that is decided by WFA on the social problem $({\sf P}_{\rm soc})$. We first introduce some protocols as sub-routines for several common distributed computation tasks via SPDZ:

\begin{enumerate}

\item $\Pi_{\min}[\llangle y \rrangle,\llangle z \rrangle]$ is a protocol that computes secretly shared output $\llangle x \rrangle$ via SPDZ, given two secretly shared values $\llangle y \rrangle,\llangle z \rrangle$, such that $\llangle x \rrangle\leftarrow \min\{\llangle y \rrangle, \llangle z \rrangle\}$. 

\item $\Pi_{<}[\llangle x \rrangle,\llangle y \rrangle]$ is a protocol that compares two secretly shared values $\llangle x \rrangle$ and $\llangle y \rrangle$ via SPDZ, and outputs 1 if $\llangle x \rrangle-\llangle y \rrangle<0$, and 0 otherwise, without revealing $x, y$.

\item $\Pi_{=}[\llangle x \rrangle,\llangle y \rrangle]$ is a protocol that compares two secretly shared values $\llangle x \rrangle$ and $\llangle y \rrangle$via SPDZ, and outputs 1 if $\llangle x \rrangle-\llangle y \rrangle=0$, and 0 otherwise, without revealing $x, y$.

\end{enumerate}
The details of these protocols can be found in Appendix~\ref{sec:protocols}. 

We divide the decision-making mechanism for group purchasing into four main stages as follows:

\vspace{-5pt}
\subsection*{\bf Stage 0: (Registration and Initialization)}

First, a group-based energy plan ${\mathfrak g}$ is registered on the blockchain. There are a group of users $X$ (such that $|X| \ge {\sf N}({\mathfrak g})$), who are interested in plan ${\mathfrak g}$. To begin the decision-making process for group purchasing, there are various initialization processes of the protocols, for instance, the pre-processing phase of SPDZ (see Appendix.~\ref{sec:append3}). After the initialization, the users will proceed to Stage 1.

\vspace{-10pt}
\subsection*{\bf Stage 1: (Data Sharing and Validation)}

We suppose that the energy bills are charged every fixed epoch. At each epoch, the users provide the energy bill data of the current epoch for the group purchasing decision-making process in a privacy-preserving manner. To generate the private input, the users secretly share their energy bill data of the current epoch $\big(\llangle a^t_i\rrangle, \llangle\beta^t_i \rrangle, \llangle\kappa^t_i \rrangle,$ $\llangle\mu_{i}^t \rrangle, \llangle \nu_{i}^t \rrangle\big)$ for the epoch $t \in [t_0, t_1]$ with others by following the procedure of the online phase in SPDZ. The users also need to validate that their secretly-shared energy bill data matches the encrypted commitments $\big({\tt Cm}(a^t_i), {\tt Cm}(\beta^t_i),$ ${\tt Cm}(\kappa^t_i), {\tt Cm}(\mu_{i}^t),$ ${\tt Cm}(\nu_{i}^t)\big)$ on the blockchain ledger  by $\Pi_{\tt dzkpCm}$, as described in Section~\ref{sec:validatebill}. Note that these commitments are prepared by the utility operators. Hence, we assume the validity of the committed values. If all secretly-shared energy bill data passes the validation of $\Pi_{\tt dzkpCm}$, the users will proceed to Stage 2.

\vspace{-10pt}
\subsection*{\bf Stage 2: (Decision-Making)}

In this stage, the users jointly compute the online decision-making algorithm (i.e., WFA on the social problem $({\sf P}_{\rm soc})$) in a privacy-preserving manner via SPDZ. It is possible that the users continue the online decision-making algorithm from the last epoch, if the last epoch's decision is not to join the group-based plan ${\mathfrak g}$. In this case, the offline optimal costs of the last epoch, $\llangle{\tt Opt}_i^{t}\rrangle$, will also be input to the online decision-making algorithm. 

The users execute WFA by a distributed protocol via SPDZ. The protocol is described by $\Pi_{\rm WFA}$ (Algorithm~\ref{alg:onlineSA}), in which the users first compute the offline optimal costs $\llangle{\tt Opt}_i^{t}\rrangle$ and $\llangle{\tt Opt}_{i,{\mathfrak g}}^{t}\rrangle$ in dynamic programming via SPDZ. Then the users will test conditions in WFA via SPDZ whether they should join energy plan ${\mathfrak g}$ or not. $\Pi_{\rm WFA}$ calls sub-routines $\Pi_{\min}, \Pi_{<}, \Pi_{=}$ to complete the tasks. 

If the decision is not to join plan ${\mathfrak g}$ at the current epoch, then the users will return to Stage 1 for the next epoch. Otherwise, the users will proceed to Stage 3, and compute the mutual compensations and carry out the payments on blockchain.

\begin{algorithm}[tbh!] 
	\caption{$\Pi_{\rm WFA}$: {\em Compute the group purchasing decision by WFA via SPDZ whether the users in $X$ should join energy plan ${\mathfrak g}$}} \label{alg:onlineSA}
{\scriptsize	
	\begin{algorithmic}[1]
		\Require $\Big(\big(\llangle a^t_i\rrangle, \llangle\beta^t_i \rrangle, \llangle\kappa^t_i \rrangle, \llangle\mu_{i}^t \rrangle, \llangle \nu_{i}^t \rrangle\big)_{t=t_0}^{t_1}\Big)_{i \in X}$,
		$\big(\llangle{\tt Opt}_i^{t_0-1}\rrangle\big)_{i \in X}$
		\Ensure {\sf Join} or {\sf Stay}
	\For {$t \in [t_0, t_1]$}
		\For {$i \in X$}
		   \State  Compute the following via SPDZ among all users:
			\begin{align}
			\llangle\kappa^t_{i,{\mathfrak g}} \rrangle \leftarrow {\tt p}_+^t({\mathfrak g}) \cdot \llangle a_i^t \rrangle \cdot\llangle\beta^t_i\rrangle + {\tt p}_{\mbox{-}}^t({\mathfrak g}) \cdot \llangle a_i^t \rrangle \cdot(1-\llangle\beta^t_i\rrangle) \notag
			\end{align}
			\State Compute the following offline optimal costs via SPDZ by calling protocol $\Pi_{\min}$:
			\begin{align}
			\llangle{\tt Opt}_i^t\rrangle & \leftarrow \Pi_{\min}\Big[\llangle{\tt Opt}_i^{t-1} \rrangle+\llangle\kappa^t_i \rrangle,\ \llangle{\tt Opt}_{i,{\mathfrak g}}^{t-1} \rrangle + \llangle\kappa^t_{i} \rrangle + \llangle\mu^t_i \rrangle + {\tt d}({\mathfrak g}) \Big] \notag\\
			\llangle{\tt Opt}_{i,{\mathfrak g}}^t\rrangle & \leftarrow \Pi_{\min}\Big[\llangle{\tt Opt}_{i,{\mathfrak g}}^{t-1}\rrangle+\llangle\kappa^t_{i,{\mathfrak g}}\rrangle,\ \llangle{\tt Opt}_i^{t-1}\rrangle +\llangle\kappa^t_{i,{\mathfrak g}}\rrangle + {\tt c}({\mathfrak g}) + \llangle\nu^t_i\rrangle \Big] \notag
			\end{align}
		\EndFor	
		\LeftComment{{\em Apply WFA to decide if switching to group plan ${\mathfrak g}$. Call $\Pi_{<}$ and $\Pi_{=}$ to test the conditions}}
		\If {$\Pi_{<}\Big[\sum_{i \in X} \llangle {\tt Opt}_{i,{\mathfrak g}}^t\rrangle +  {\tt c}({\mathfrak g}) + \llangle\nu^t_i\rrangle, \sum_{i \in X} \llangle{\tt Opt}_i^t\rrangle\Big]=1$ and\\ \qquad $\Pi_{=}\Big[\sum_{i \in X} \llangle{\tt Opt}_{i,{\mathfrak g}}^t \rrangle, \sum_{i \in X} \llangle{\tt Opt}_{i,{\mathfrak g}}^{t-1}\rrangle+\llangle\kappa^t_{i,{\mathfrak g}}\rrangle\Big]=1$} 
		\State \Return {\sf Join}
		\EndIf
	\EndFor				
	\State \Return {\sf Stay}
	\end{algorithmic}	
}	
\end{algorithm}

\begin{algorithm}[tbh!] 
	\caption{$\Pi_{\rm MC}$: {\em Compute compensated switching cost $\big(\llangle \theta_i \rrangle\big)_{i \in X}$ according to either egalitarian cost-sharing or proportional cost-sharing and return the net mutual compensations}} \label{alg:MC}
{\scriptsize	
	\begin{algorithmic}[1]
		\Require 
		$\big(\llangle \nu_{i}^t \rrangle, \llangle{\tt Opt}_{i}^t\rrangle, \llangle{\tt Opt}_{i,{\mathfrak g}}^t\rrangle\big)_{i \in X}$
		\Ensure $\big(\llangle \theta_i \rrangle\big)_{i \in X}$
		\For {$i \in X$}
		   \State  Compute the following via SPDZ among all users:
		   \begin{align}
		   \llangle C^t_{i}({\mathcal E} \backslash \{ {\mathfrak g} \})\rrangle & \leftarrow  \llangle{\tt Opt}_i^t \rrangle \\
		   \llangle C^t_{i}({\mathfrak g}) & \leftarrow  \llangle {\tt Opt}_{i,{\mathfrak g}}^t\rrangle +  {\tt c}({\mathfrak g}) + \llangle\nu^t_i\rrangle	   
		   \end{align}
		\EndFor		
		   \State Compute and reveal $\sum_{i \in X} \llangle C^t_{i}({\mathcal E} \backslash \{ {\mathfrak g} \})\rrangle$, $\sum_{i \in X} \llangle C^t_{i}({\mathfrak g})\rrangle$, and their MACs to all users.
		\If{Checking MACs failed}
		\State Abort
		\EndIf
		\For {$i \in X$}
		\State  Compute the following via SPDZ among all users:		
			\begin{align}
			\llangle\theta_i \rrangle \leftarrow
			\left\{
			\begin{array}{@{}l@{}l@{}}
			 \llangle C^t_i({\mathcal E} \backslash \{ {\mathfrak g} \})\rrangle  +  \frac{ \sum_{i' \in X}  C^t_{i'}({\mathfrak g}) -   \sum_{i' \in X} C^t_{i'}({\mathcal E} \backslash \{ {\mathfrak g} \})  }{|X|} -  \llangle {\tt Opt}_{i,{\mathfrak g}}^t \rrangle, \ \ & \mbox{for ega. cost-sharing}\\
			 \llangle C^t_i({\mathcal E} \backslash \{ {\mathfrak g} \})\rrangle \cdot \frac{\sum_{i' \in X} C^t_{i'}({\mathfrak g})}{\sum_{i' \in X} C^t_{i'}({\mathcal E} \backslash \{ {\mathfrak g} \}) } -  \llangle {\tt Opt}_i^{t}[{\mathfrak g}] \rrangle, & \mbox{for prop. cost-sharing}
			\end{array}
			\right. \notag
			\end{align}
		\EndFor				
		\LeftComment{{\em Compute the net mutual compensations via SPDZ}}		
		\For {$i \in X$}
			\State $\llangle \phi_i \rrangle \leftarrow \llangle \theta_i \rrangle - {\tt c}({\mathfrak g}) - \llangle\nu^t_i\rrangle$
			\State Reveal $\llangle \phi_i \rrangle$ and its MAC to user $i$ only
			\If{Checking MAC failed}
			\State Abort
			\EndIf
		\EndFor				
	\end{algorithmic}	 
}	
\end{algorithm}

\vspace{-10pt}
\subsection*{\bf Stage 3: (Mutual Compensations)}

After deciding to join plan ${\mathfrak g}$, the users compute their mutual compensations in a privacy-preserving manner via SPDZ based on the secretly-shared input from Stage 2. They will pay the mutual compensations on the blockchain ledger afterwards and the decision will be recorded on the ledger for the energy provider of plan ${\mathfrak g}$.

The users compute mutual compensations by a distributed protocol via SPDZ. The protocol is described by $\Pi_{\rm MC}$ (Algorithm~\ref{alg:MC}), in which the users compute the compensated switching cost $\big(\llangle \theta_i \rrangle\big)_{i \in X}$ by Theorem~\ref{thm:payments}, and the net mutual compensations ($\phi_i$) between the compensated and original switching costs, defined by $\phi_i \triangleq \theta_i - {\tt Cost}_{\tt Sw}^t[{x_i^{t-1}},{\mathfrak g}]$. In this stage,  $\big(\llangle \phi_i \rrangle\big)_{i \in X}$ will remain secretly-shared, and will be used in Stage 4 for payments on blockchain.

\vspace{-10pt}
\subsection*{\bf Stage 4: (Payments)}

The users pay the net mutual compensations $\big(\phi_i \big)_{i \in X}$ by ERC20 tokens on the blockchain. The users generate a multi-transaction request with concealed transaction values $\big( {\tt Cm}(\phi_i) \big)_{i \in X}$. The transaction generation will be handled in a privacy-preserving manner via SPDZ, which follows a similar privacy-preserving payment system in \cite{CWZ21blockchain}. Because of the paucity of space, we skip the protocol, and defer the details to Appendix~\ref{sec:payments}.

%% file: experiment.tex
\begin{figure*}[t]
\centering  
\includegraphics[width=0.85\linewidth]{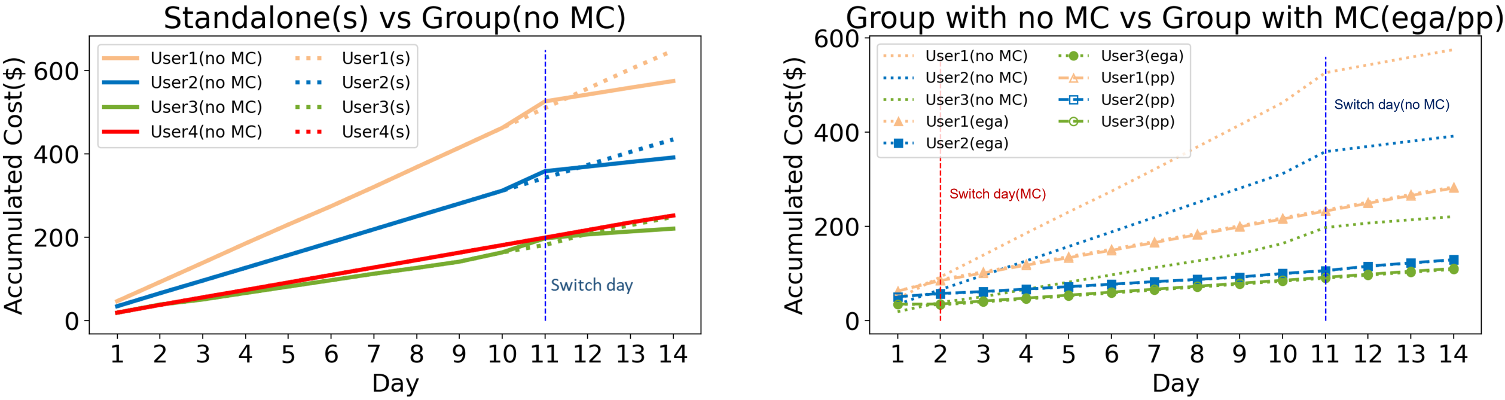} 
\vspace{-10pt}
\caption{Evaluation of online decision algorithm, comparing the online solutions with and without mutual compensations.} \label{fig:online}  \vspace{-5pt}
\end{figure*}

\begin{figure*}[t]
\centering  
\includegraphics[width=0.85\linewidth]{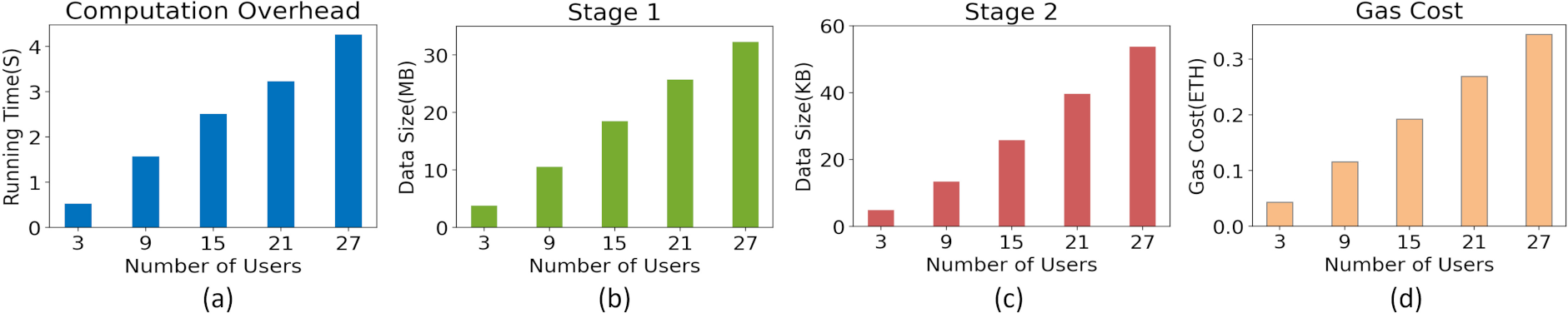} 
\vspace{-10pt}
\caption{SPDZ performance and smart contract gas costs } \label{fig:SPDZ}  \vspace{-5pt}
\end{figure*}

\section{Evaluation} \label{sec:eval}

This section presents an empirical evaluation of our solution, including the online decision-making algorithm, SPDZ performance and the gas costs of smart contract on Ethereum.

\subsection{Online Decision-Making Algorithm}

\begin{figure}[!h]
\centering  \vspace{-5pt}
\includegraphics[width=1\linewidth]{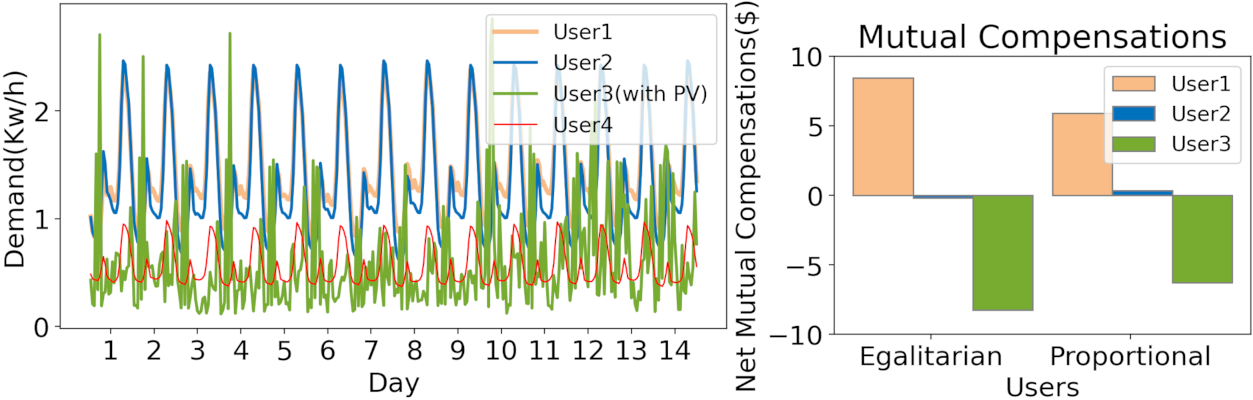}  \vspace{-5pt}
\caption{Data trace of energy consumption and net mutual compensations.} \vspace{-10pt}
\label{fig:trace_mc} 
\end{figure}

We present an evaluation of our online algorithm using real-world hourly electricity consumption data of 4 users (see Figure \ref{fig:trace_mc} and User 3 has a solar PV that can
probably export extra electricity to the grid). The data trace represents the total daily electricity usage of each household in a period of consecutive 14 days. The energy plan tariffs in table \ref{tbl:cost} are characterized by peak hour rates, off-peak hour rates and disconnection fees. The peak hour period is from 8am to 8pm, and the off-peak hour period is from 9pm to 7am respectively. We set a zero connection fee in our experiments. We evaluated the performance of the online decision algorithm without mutual compensations. In Figure \ref{fig:online}, we observe a small spike in the accumulated cost on Day 11 among Users 1, 2 and 3 due to switching to the group plan and paying off the disconnection fee of the standard plan. But the users enjoy more cost-saving afterwards. We also study different mutual compensation schemes. We observe that User 1 pays more with egalitarian cost-sharing (ega) to compensate user 2 and user 3, whereas Users 1 and 2 pay more with proportional cost-sharing (pp) to compensate user 3 in the beginning, but all of them will be benefited considerably in a long term. At last, we study the performance of our online algorithm considering with and without mutual compensations (MC). We observe that with mutual compensations (MC), all users join the group-based plan earlier on Day 2. Hence, the energy costs of all users are lower than those without mutual compensations. All users are benefited substantially with more than $50\%$ cost reduction, when joining the group-based plan earlier with mutual compensations.

\begin{table}[!h]
    \centering
    \caption{\label{tbl:cost}Table of Energy Plan Tariffs.} 
    \begin{tabularx}{0.95\linewidth}{@{}c | c | c | c@{}}
    \hline  \hline   
      & Peak (\$) & Off-Peak (\$) & Disconnect Fee (\$) \\ \hline \hline
    Standalone Plan & 1.6 & 1.0 & 16 \\ \hline
  Group Plan & 0.6 & 0.3 & 30 \\ 
    \hline  \hline    
    \end{tabularx}
\end{table}

\subsection{SPDZ Performance}

We evaluated the performance of SPDZ in Stage 2. We do not show the performance of Stage 3 and Stage 4, because of their small overhead. We consider a single fortnight period and up to 27 users.

{\bf Computation Running Time}: All the results are averaged over 10 instances. Figure \ref{fig:SPDZ} represents the average running time of each user in Stage 2. We observe that the running time only increases linearly with the number of users with moderate overhead.

{\bf Communication Overhead}: We evaluated the communication data size among Stage 1 and Stage 2. Figure \ref{fig:SPDZ} shows that the total communication data size grows linearly with the number of users in both Stage 1 and Stage 2 with moderate overhead.

\vspace{-5pt}
\subsection{Smart Contract Gas Costs}
We implemented the payment system of mutual compensations by a smart contract on Ethereum using Solidity programming language. We measured the gas costs of our smart contract, it represents the amount of computational resources needed to execute a single transaction. The gas price is set by the operator, and most miners will choose the smart contract with a higher gas price in a block. We use the standard gas price 54 Gwei to simulate the transaction cost in Ether. Figure \ref{fig:SPDZ} shows the gas costs. We observe that the gas cost increases linearly, starting from 0.0437 ETH with 3 users to 0.3445 ETH with 27 users. Overall, our smart contract generated comparable gas costs as other privacy-preserving smart contract implementations in the literature.

\iffalse

\begin{table}[!h]
    \centering
    \caption{\label{tbl:cost}Table of Gas Costs.} 
    \begin{tabularx}{\linewidth}{c | c | c | c}
    \hline  \hline   
      & Gas Cost & Ether & Input Size{\tiny (bytes)} \\ \hline \hline
    submit() & 98K & 0.0053 & 64 + 64  $\times{N}$ \\ \hline
  confirm() & 3418K & 0.184 & 128 + 180  $\times{N_b}$ \\ 
    \hline  \hline    
    \end{tabularx}
\end{table}

\fi

%% file: related.tex
\section{Related Work} \label{sec:related}
In this section, we present the related work in the literature. 

The problem of energy plan selection has been studied as an online decision problem, which belongs to the class of Metrical Task System (MTS) problem \cite{BEY05online} and online convex optimization (OCO) problem with switching cost \cite{oco}. For a general setting with $n$ discrete states, the MTS problem is known to have a competitive ratio of $2n-1$ \cite{mts}. An instance of MTS is the energy generation scheduling in microgrids \cite{chase,rchase,pchase}. \cite{lcp,disc} proposed online algorithms for OCO. The problem of energy plan selection is also studied as an MTS problem in \cite{ZCC19energyplan}. Most of the extant literature of online decision problems does not consider the settings of cooperative multi-players, in which the players may decide their joint online decisions.  Recently, the work in \cite{ZC21energyplan} considers cooperative multiplayer online decision problems of group purchasing energy plan selection problem. This paper extends the online algorithm with mutual compensations in \cite{ZC21energyplan} by considering a different mutual compensation scheme and the possibility of energy export.

Blockchain technology has been applied to diverse aspects of energy systems. Among the applications, \cite{ggk19block} applied blockchain to mitigate trust in peer-to-peer electric vehicle charging. Blockchain has been applied to microgrid energy exchange and wholesale markets by prosumers \cite{m18brooklyn}. Renewable energy credits and emissions trading are also applications of blockchain \cite{kbue20gecko}. In these applications, the goal of blockchain is to improve transparency and reduce settlement times, since blockchain can ensure integrity and consistency of transactions and settlement on an open ledger. See a recent survey about blockchain applications to energy systems \cite{a19survey}. However, very few studies considered the privacy of blockchain, even transaction data on the ledger is entirely disclosed to the public. \cite{CWZ21blockchain} recently proposed a privacy-preserving solution for energy storage sharing using blockchain and secure multi-party computation. This work draws on similar concepts from privacy-preserving blockchain, but also addressing a different problem of decentralized group purchasing of energy plans. It is worth noting that supporting privacy on blockchain is a crucial research topic in cryptography and security. There are several privacy-preserving blockchain platforms with support of privacy (e.g., ZCash, Monero, Zether, Tornado Cash \cite{zcash,monero,bunz20zether}). 

There are two major approaches in privacy-preserving techniques in the literature: (1) data obfuscation that masks private data with random noise, (2) secure multi-party computation that hides private data while allowing the data to be computed confidentially. Differential privacy \cite{dwork2006calibrating}, a main example of data obfuscation, is often used in privacy-preserving data mining in a very large dataset \cite{ZJ14smeter}. Note that there is an intrinsic trade-off between accuracy and privacy in differential privacy. On the other hand, secure multi-party computation \cite{goldreich1998secure, du2001secure} traditionally employed garbled circuits \cite{HazayL10}, which  have a high computational complexity, and homomorphic cryptosystems \cite{cramer2001multiparty,lcwz20privsharing,RL19EV}, which need a trusted setup for key generation. Recently, information-theoretical secret-sharing \cite{cramer2015secure, dklpss13spdz} has been utilized for secure multi-party computation, which provides high efficiency and requires no trusted third-party setup.  Particularly, SPDZ \cite{cramer2015secure} is an efficient MPC protocol based on secret-sharing and can safeguard against a dishonest majority of users.

%% file: concl.tex
\vspace{-5pt}
\section{Conclusion} \label{sec:concl}

The emergence of competitive retail energy markets introduces group purchasing of energy plans. However, traditional group purchasing mediated by a trusted third-party suffers from the lack of privacy and transparency. In this paper, we introduced a novel paradigm of {\em decentralized privacy-preserving} group purchasing. We combine competitive online algorithms, secure multi-party computations and zero-knowledge on blockchain as a holistic solution to enable users to make coordinated switch decisions in a decentralized manner, without a trusted third-party. We implemented our decentralized group purchasing solution by multi-party computation protocol (SPDZ) and a smart contract on Solidity-supported blockchain platform. 
Although we implemented it as a smart contract on permissionless blockchain, it can be implemented on a permissioned blockchain, on which the gas cost is not a concern. In future work, we will extend our online decision algorithms to complex scenarios (e.g., with termination penalty in group purchasing).

%% file: append1.tex
\section{Proofs of Results} \label{sec:append1}

\begin{customthm}{2}
If the connection fees and disconnection fees of all energy plans are identical, i.e., ${\tt c}_{j} = {\tt c}$ and ${\tt d}_{j} = {\tt d}$ for all $j \in {\mathcal E}$, then 
\[
{\tt Cost}_i[\hat{x}_{i}] \le 3 \cdot {\tt Opt}^{T}_i[{x^\ast}^T_i] 
\]	
\end{customthm}
\begin{proof}
Since ${\tt c}_{j} = {\tt c}$ and ${\tt d}_{j} = {\tt d}$ for all $j \in {\mathcal E}$, the switching cost is a constant between all states. In this case, we can reduce the problem of $n$ states to the one of 2 states only. When deciding the switching decision, one only needs to consider whether to stay with the current stay or change to another state that has a lower state cost (as the switching cost of any state is a constant).
\end{proof}

\begin{customthm}{3}
We consider two schemes of mutual compensations:
\begin{enumerate}

\item{\bf Egalitarian Cost-sharing:}
\begin{align} 
\theta_i  =\  & C^t_i({\mathcal E} \backslash \{ {\mathfrak g} \})  +  \frac{ \sum_{i' \in X} \big( C^t_{i'}({\mathfrak g}) - C^t_{i'}({\mathcal E} \backslash \{ {\mathfrak g} \})  \big)}{|X|} - {\tt Opt}_i^{t}[{\mathfrak g}] 
\end{align}

\item{\bf Proportional Cost-sharing:}
\begin{align} 
\theta_i  =\  & C^t_i({\mathcal E} \backslash \{ {\mathfrak g} \}) \frac{\sum_{i' \in X} C^t_{i'}({\mathfrak g})}{\sum_{i' \in X} C^t_{i'}({\mathcal E} \backslash \{ {\mathfrak g} \}) } - {\tt Opt}_i^{t}[{\mathfrak g}] 
\end{align}

\end{enumerate}

Then $(\theta_i )_{i \in X}$ in both schemes will satisfy individual rationality and budget balance, when group feasibility is satisfied.
\end{customthm}

\begin{proof}
First, we consider egalitarian cost-sharing. Note that ${\tt Opt}_i^{t}[{\mathfrak g}] + \theta_i  \ge 0$ for all $i \in X$.
Suppose group feasibility is satisfied, then we obtain $\sum_{i \in X} \big( C^t_i({\mathfrak g}) - C^t_i({\mathcal E} \backslash \{ {\mathfrak g} \})  \big) < 0$. We show individual rationality as follows:
\begin{align}
{\tt Opt}_i^{t}[{\mathfrak g}] + \theta_i = C^t_i({\mathcal E} \backslash \{ {\mathfrak g} \})  +  \frac{ \sum_{i' \in X} \big( C^t_{i'}({\mathfrak g}) - C^t_{i'}({\mathcal E} \backslash \{ {\mathfrak g} \})  \big)}{|X|} < C^t_i({\mathcal E} \backslash \{ {\mathfrak g} \})
\notag 
\end{align} 

Next, we show budget balance as follows:
\begin{align}
& \sum_{i \in X} \Big( {\tt Opt}_i^{t}[{\mathfrak g}] + \theta_i \Big) = \sum_{i \in X} \Big( C^t_i({\mathcal E} \backslash \{ {\mathfrak g} \})  +  \frac{  \sum_{i \in X} \big( C^t_i({\mathfrak g}) - C^t_i({\mathcal E} \backslash \{ {\mathfrak g} \})  \big)  }{|X|} \Big) \notag \\
= &  \sum_{i \in X} C^t_i({\mathcal E} \backslash \{ {\mathfrak g} \})  +  \sum_{i \in X} C^t_i({\mathfrak g}) - \sum_{i \in X} C^t_i({\mathcal E} \backslash \{ {\mathfrak g} \}) = \sum_{i \in X} C^t_i({\mathfrak g})  \notag 
\end{align} 

Similarly, we consider proportional cost-sharing. Suppose group feasibility is satisfied, then we obtain $\frac{\sum_{i \in X} C^t_i({\mathfrak g})}{\sum_{i \in X} C^t_i({\mathcal E} \backslash \{ {\mathfrak g} \}) } < 1$. We show individual rationality as follows:
\begin{align}
{\tt Opt}_i^{t}[{\mathfrak g}] + \theta_i = C^t_i({\mathcal E} \backslash \{ {\mathfrak g} \}) \frac{\sum_{i' \in X} C^t_{i'}({\mathfrak g})}{\sum_{i' \in X} C^t_{i'}({\mathcal E} \backslash \{ {\mathfrak g} \}) } < C^t_{i'}({\mathcal E} \backslash \{ {\mathfrak g} \})
\notag 
\end{align} 

Next, we show budget balance as follows:
\begin{align}
& \sum_{i \in X} \Big( {\tt Opt}_i^{t}[{\mathfrak g}] + \theta_i \Big) = \sum_{i \in X} \Big( C^t_i({\mathcal E} \backslash \{ {\mathfrak g} \}) \frac{\sum_{i' \in X} C^t_{i'}({\mathfrak g})}{\sum_{i' \in X} C^t_{i'}({\mathcal E} \backslash \{ {\mathfrak g} \}) } \Big)  = \sum_{i \in X} C^t_i({\mathfrak g})  \notag 
\end{align} 
\end{proof}

\section{Zero-Knowledge Proofs} \label{sec:append2}

In this section, we provide a brief explanation of the concepts of Zero-Knowledge Proofs (ZKPs) in this section. More details can be found in a standard cryptography textbook (e.g., \cite{crytobk}). 

First, we define some system parameters. Denote by ${\mathbb Z}_p = \{0, ..., p-1\}$ a finite field of integers modulo $p$, for encrypting private data. For brevity, we simply write ``$x+y$'' and ``$x\cdot y$'' for modular arithmetic without explicitly mentioning``${\tt mod\ } p$''. We consider a usual finite group ${\mathbb G}$ of order $p$. We pick $g, h$ as two generators of ${\mathbb G}$, such that they can generate every element in ${\mathbb G}$ by taking proper powers, namely, for each $e \in {\mathbb G}$, there exist $x,y \in {\mathbb Z}_p$ such that $e = g^x = h^y$. The classical discrete logarithmic assumption states that given $g^x$, it is computationally hard to obtain $x$, which underlies the security of many cryptosystems.

In a zero-knowledge proof (ZKP) (of knowledge), a prover convinces a verifier of the knowledge of a secret without revealing the secret. For example, to show the knowledge of $(x, {\tt r})$ for ${\tt Cm}(x, {\tt r})$ without revealing $(x, {\tt r})$. A zero-knowledge proof of knowledge should satisfy completeness (i.e., the prover always can convince the verifier if knowing the secret), soundness (i.e., the prover cannot convince a verifier if not knowing the secret) and zero-knowledge (i.e., the verifier cannot learn the secret).

\subsection{$\Sigma$-Protocol}

$\Sigma$-Protocol is a general approach to construct zero-knowledge proofs. Given a computationally non-invertible function $f(\cdot)$ that satisfies homomorphic property $f(a+b) = f(a) + f(b)$ and $f(x)=y$, one can prove the knowledge of the concealed $x$:

\begin{enumerate}

\item First, the prover sends a commitment $y' = f(x')$, for a random $x'$, to the verifier. 

\item Next, the verifier replies with a random challenge $\beta$.

\item The prover replies with $z = x' + \beta \cdot x$ \mbox{(which does not reveal $x$)}. 

\item Finally, the verifier checks whether $f(z) \overset{?}{=} y' + \beta \cdot y$. 

\end{enumerate}

\subsection{$\Sigma$-Protocol Based Zero-knowledge Proofs}

Next, we present several instances of zero-knowledge proofs based on $\Sigma$-protocol:

\begin{itemize}

\item {\bf ZKP of Commitment (${\tt zkpCm}[x, {\tt Cm}(x)]$)}: Given ${\tt Cm}(x, r)$, a prover can convince a verifier of the knowledge of $x$ without revealing $(x, r)$. The corresponding protocol is described by $\Pi_{\tt zkpCm}$.

\item {\bf ZKP of Membership (${\tt zkpMbs}[x_i$$\in$${\mathcal X}]$)}: Given a set ${\mathcal X} = \{x_1, ..., x_n \}$ and ${\tt Cm}(x, {\tt r})$, a prover can convince a verifier of the knowledge of $x \in {\mathcal X}$ without revealing $x$. The corresponding protocol is described by $\Pi_{\tt zkpMbs}$.

\item {\bf ZKP of Non-Negativity (${\tt zkpNN}[x$$\ge$$0]$)}: Given ${\tt Cm}(x, {\tt r})$, a prover can convince a verifier of the knowledge of $x \ge 0$ without revealing $x$. The corresponding protocol is described by $\Pi_{\tt zkpNN}$.

\end{itemize}
%The detailed constructions of these zero-knowledge proofs can be found in Appendix.~\ref{sec:append2}.

\begin{algorithm}[hbt!] 
	\caption{$\Pi_{\tt zkpCm}$: {\em Prove the knowledge of $(x, {\tt r})$ in a given commitment ${\tt Cm}(x, {\tt r})$, without revealing $(x, {\tt r})$}} \label{alg:zkpCm}
	{\scriptsize
	\begin{algorithmic}[1]		
		\Require ${\tt Cm}(x, {\tt r})$ (known to the verifier and prover); $(x,{\tt r})$ (known to the prover only)
		\Ensure {\sf Pass} or {\sf Fail}
			
		\State 
		The prover generates a pair of random numbers $(x', {\tt r}') \overset{\$}{\leftarrow} {\mathbb Z}^2_p$ and announces the commitment ${\tt Cm}(x', {\tt r}')$ to the verifier
		
		\State 
		The verifier generates a random number $\psi \overset{\$}{\leftarrow} {\mathbb Z}_p$ and announces $\psi$ to the prover

		\State 
		The prover computes $z_x \leftarrow x' + \psi \cdot x$ and $z_{\tt r} \leftarrow {\tt r}' + \psi \cdot {\tt r}$, and announces $(z_x, z_{\tt r})$ to the verifier

		\Statex \LeftComment{{\em The verifier checks the following}}
		\If{$g^{z_x} \cdot h^{z_{\tt r}} =  {\tt Cm}(x', {\tt r}') \cdot {\tt Cm}(x, {\tt r})^\psi$} 
		
		\State \Return {\sf Pass}

		\Else

		\State \Return {\sf Fail}
		
		\EndIf
		
		\Statex \LeftComment{{\em The ZKP of commitment is ${\sf zkpCm}[x, {\tt Cm}(x)] = \big\{ {\tt Cm}(x, {\tt r}); {\tt Cm}(x', {\tt r}'), z_x, z_{\tt r}\big\}$}}

	\end{algorithmic}	
}
\end{algorithm}

\begin{algorithm}[hbt!] 
	\caption{$\Pi_{\tt zkpMbs}$: {\em Prove the knowledge that $x_i \in {\mathcal X} \triangleq \{x_1,$ ..., $x_n\}$, given commitment ${\tt Cm}(x_i, {\tt r})$, without revealing~$x_i$}} \label{alg:zkpMbs}
	{\scriptsize
	\begin{algorithmic}[1]		
		\Require $ {\mathcal X}, {\tt Cm}(x_i, {\tt r})$ (known to the verifier and prover); $(x_i,{\tt r})$ (known to the prover only)
		\Ensure {\sf Pass} or {\sf Fail}
			
		\State 
		The prover generates a pair of random numbers $(x'_j, {\tt r}'_j) \overset{\$}{\leftarrow} {\mathbb Z}^2_p$ and announces the commitments ${\tt Cm}(x'_j, {\tt r}'_j)$ for all $j\in\{1,...,n\}$ to the verifier
		
		\State 
		The prover generates a random number 
		$\psi_j \overset{\$}{\leftarrow} {\mathbb Z}_p$ for each $j\in\{1,...,n\}\backslash\{i\}$, and computes
		\[
		z_{x_j} \leftarrow
		\begin{cases}
		x'_j + \psi_j \cdot (x_i - x_j) ,& \mbox{if\ } j\in\{1,...,n\}\backslash\{i\}\\
		x'_i, & \mbox{if\ } j = i\\
		\end{cases}
		\] 
		
		\State 
		The verifier generates a random number $\psi \overset{\$}{\leftarrow} {\mathbb Z}_p$ and announces $\psi$ to the prover

		\State 
		The prover sets $\psi_i \leftarrow \psi - \sum_{j\ne i}\psi_j$ and $z_{{\tt r}_j} \leftarrow {\tt r}'_j + {\tt r} \cdot \psi_j$ for all $j\in\{1,...,n\}$, and then announces $(\psi_j, z_{{\tt r}_j})_{j=1}^n$ to the verifier

		\Statex \LeftComment{{\em The verifier checks the following}}
		\If{$g^{z_{x_j}} \cdot h^{z_{{\tt r}_j}} = {\tt Cm}(x'_j, {\tt r}'_j) \cdot \Big( \frac{{\tt Cm}(x_i, {\tt r})}{g^{x_j}} \Big)^{\psi_j}$ for all $j\in\{1,...,n\}$ and $\psi = \sum_{i=1}^n \psi_j$} 
		
		\State \Return {\sf Pass}

		\Else

		\State \Return {\sf Fail}
		
		\EndIf
		
		\Statex \LeftComment{\mbox{\em The ZKP of membership is ${\sf zkpMbs}[x_i$$\in$${\mathcal X}] = \big\{ {\tt Cm}(x_i, {\tt r}); \{{\tt Cm}(x'_j, {\tt r}'_j)\}_{j=1}^n, (\psi_j, z_{x_j}, z_{{\tt r}_j})_{j=1}^n\big\}$}}

	\end{algorithmic}	
	}
\end{algorithm}

\begin{algorithm}[hbt!] 
	\caption{$\Pi_{\tt zkpNN}$: {\em Prove the knowledge that $x \ge 0$ by showing there exist $(b_1, ..., b_m)$ such that $b_i \in \{0, 1\}$ for $i \in \{0, ..., m\}$ and $\sum_{i=1}^{m}b_i \cdot 2^{i-1} = x$, given commitment ${\tt Cm}(x, {\tt r})$, without revealing $x$}} \label{alg:zkpNN}
	{\scriptsize
	\begin{algorithmic}[1]		
		\Require ${\tt Cm}(x, {\tt r})$ (known to the verifier and prover); $(x,{\tt r})$ (known to the prover only)
		\Ensure {\sf Pass} or {\sf Fail}
			
		\State The prover announces $({\tt Cm}(b_i, {\tt r}_i))_{i=1}^{m}$ to the verifier, and uses $\Pi_{\tt zkpMbs}[\{0,1\}, {\tt Cm}(b_i, {\tt r}_i), b_i]$ to prove that $b_i \in \{0, 1\}$	
			
		\State 
		The prover generates a random number ${\tt r}' \overset{\$}{\leftarrow} {\mathbb Z}_p$ and announces the commitment ${\tt Cm}(0, {\tt r}')$ to the verifier
		
		\State 
		The verifier generates a random number $\psi \overset{\$}{\leftarrow} {\mathbb Z}_p$ and announces $\psi$ to the prover
		
		\State 
		The prover computes $z_{\tt r} \leftarrow {\tt r}' + \psi \cdot (\sum_{i=1}^m {\tt r}_i \cdot 2^{i-1} - {\tt r})$ and announces $z_{\tt r}$ to the verifier

		\Statex \LeftComment{{\em The verifier checks the following}}
		\If{$h^{z_{\tt r}} = {\tt Cm}(0,{\tt r}') \cdot {\tt Cm}(x, {\tt r})^{-\psi}\cdot\prod_{i=1}^m {\tt Cm}(b_i, {\tt r}_i)^{\psi\cdot 2^{i-1}}$} 
		
		\State \Return {\sf Pass}

		\Else

		\State \Return {\sf Fail}
		
		\EndIf
		
		\Statex \LeftComment{{\em The ZKP of non-negativity is ${\sf zkpNN}[x$$\ge$$0]= \big\{ {\tt Cm}(x, {\tt r}), \{{\tt Cm}(b_i, {\tt r}_i)\}_{i=1}^m; {\tt Cm}(0, {\tt r}'), z_{\tt r}\big\}$}}

	\end{algorithmic}	
	}
\end{algorithm}

\subsection{Non-interactive Zero-knowledge Proofs}

An interactive zero-knowledge proof that requires  a verifier-provided challenge can be converted to a non-interactive one by Fiat-Shamir heuristic to remove the verifier-provided challenge. 

Let ${\mathcal H}(\cdot)\mapsto {\mathbb Z}_p$ be a cryptographic hash function. Given a list of commitments (${\tt Cm}_1, ..., {\tt Cm}_r$), one can map to a single hash value by  ${\mathcal H}({\tt Cm}_1|...|{\tt Cm}_r)$, where the input is the concatenated string of (${\tt Cm}_1, ..., {\tt Cm}_r$). In a $\Sigma$-protocol, one can set the challenge by $\beta ={\mathcal H}({\tt Cm}_1|$$...|{\tt Cm}_r)$, where (${\tt Cm}_1, ..., {\tt Cm}_r$) are all the commitments generated by the prover prior to the step of verifier-provided challenge (Step 2 of $\Sigma$-protocol). Hence, the prover does not wait for the verifier-provided random challenge, and instead generates the random challenge himself. The verifier will generate the same challenge following the same procedure for verification. We denote the non-interactive versions of the previous zero-knowledge proofs by {\tt nzkpCm}, {\tt nzkpSum}, {\tt nzkpMbs}, {\tt nzkpNN}, respectively.

%% file: append3.tex
\section{SPDZ Protocol} \label{sec:append3}

In this section, we present a simplified version of SPDZ protocol for the clarity of exposition. The full version can be found in \cite{cramer2015secure, dklpss13spdz}. 

There are three phases in SPDZ protocol: (1) pre-processing phase, (2) online phase, and (3) output and validation phase. We write $\langle x \rangle$ as a {\em secretly shared} number, meaning that there is a vector $(x_1, ..., x_n)$, such that each party $i$ knows only $x_i$. To reveal the secretly shared number $\langle x \rangle$, each party $i$ broadcasts $x_i$ to other parties. Then each party can reconstruct $x = \sum_{i=1}^n x_i$. We write $\llangle  x \rrangle$ meaning that both $\langle  x \rangle$ and the respective MAC $\langle \gamma(x) \rangle$ are secretly shared.

\subsection{Online Phase}

In the online phase, the parties can jointly compute an arithmetic circuit, consisting of additions and multiplications with secretly shared input numbers.

\subsubsection{Addition}
\

Given secretly shared $\langle x \rangle$ and $\langle y \rangle$, and a public known constant $c$, the following operations can be attained by local computation at each party, and then the outcome can be assembled from the individual shares: 
\begin{enumerate}

\item[{\tt A1})] $\langle x \rangle + \langle y \rangle$ can be computed by $(x_1 + y_1, ..., x_n + y_n)$.

\item[{\tt A2})] $c \cdot \langle x \rangle$ can be computed by $(c \cdot x_1, ..., c \cdot x_n)$.

\item[{\tt A3})] $c + \langle x \rangle$ can be computed by $(c + x_1, x_2, ..., x_n)$.

\end{enumerate}

\subsubsection{Multiplication}
\

Given secretly shared $\langle x \rangle$ and $\langle y \rangle$, computing the product  $\langle x \rangle \cdot \langle y \rangle$ involves a given multiplication triple. A multiplication triple is defined by $(\langle a \rangle, \langle b \rangle, \langle c \rangle)$, where $a, b$ are some unknown random numbers and $c = a \cdot b$, are three secretly shared numbers already distributed among the parties. The triple is assumed to be prepared in a pre-processing phase. To compute  $\langle x  \rangle \cdot \langle  y \rangle$, it follows the below steps of operations ({\tt A4}):
\begin{enumerate}

\item[{\tt A4.1})] Compute $\langle\epsilon \rangle = \langle x  \rangle - \langle  a \rangle$ (by {\tt A1}). Then, reveal $\langle\epsilon \rangle$, which does not reveal $x$.

\item[{\tt A4.2})] Compute $\langle \delta \rangle = \langle y   \rangle - \langle  b \rangle$. Then, reveal $\langle\delta \rangle$.

\item[{\tt A4.3})] Finally, compute $\langle x  \rangle \cdot \langle  y \rangle = \langle c \rangle  + \epsilon  \cdot  \langle b \rangle +  \delta  \cdot  \langle a \rangle + \epsilon  \cdot  \delta$ (by {\tt A1}-{\tt A3}).

\end{enumerate}

\subsubsection{Message Authentication Code}
\

To safeguard against dishonest parties, who may perform incorrect computation, an information-theoretical message authentication code (MAC) can be used for verification. We write a MAC key as a global number $\widetilde{\alpha}$, which is unknown to the parties, and is secretly shared as $\langle \widetilde{\alpha} \rangle$. Every secretly shared number is encoded by a MAC as $\gamma(x) = \widetilde{\alpha} x$, which is secretly shared as $\langle \gamma(x) \rangle$. For each $\langle x \rangle$, each party $i$ holds a tuple $(x_i, \gamma(x)_i)$ and $\widetilde{\alpha}_i$, where $x = \sum_{i=1}^n x_i$, $\widetilde{\alpha} = \sum_{i=1}^n \widetilde{\alpha}_i$ and $\gamma(x) = \widetilde{\alpha} x = \sum_{i=1}^n  \gamma(x)_i$. If any party tries to modify his share $x_i$ uncoordinatedly, then he also needs to modify $\gamma(x)_i$ accordingly. Otherwise, $\gamma(x)$ will be inconsistent. However, it is difficult to modify $\gamma(x)_i$ without coordination among the parties, such that $\widetilde{\alpha} x = \sum_{i=1}^n  \gamma(x)_i$. Hence, it is possible to detect incorrect computation (possibly by dishonest parties) by checking the MAC.

To check the consistency of $x$, there is no need to reveal $\langle \widetilde{\alpha} \rangle$. One only needs to reveal $\langle x \rangle$, and then reveals $\widetilde{\alpha}_i - x \cdot \gamma(x)_i$ from each party $i$. One can check whether $\sum_{i=1}^n (\widetilde{\alpha}_i - x \cdot \gamma(x)_i) \overset{?}{=} 0$ for consistency. To prevent a dishonest party from modifying his share $x_i$ after learning the other party's $x_j$. Each party needs to commit his share $x_i$ before revealing $x_i$ to others.

To maintain the consistency of MAC for operations {\tt A1}-{\tt A4}, the MAC needs to be updated accordingly as follows:
\begin{enumerate}

\item[{\tt B1})] $\langle x \rangle + \langle y \rangle$: Update MAC by $(\gamma(x)_1 + \gamma(y)_1, ..., \gamma(x)_n + \gamma(y)_n)$.

\item[{\tt B2})] $c \cdot \langle x \rangle$: Update MAC by $(c \cdot \gamma(x)_1, ..., c \cdot \gamma(x)_n)$.

\item[{\tt B3})] $c + \langle x \rangle$: Update MAC by $(c\cdot\alpha_1 + \gamma(x)_1, ..., c\cdot\alpha_n + \gamma(x)_n)$.

\item[{\tt B4})] $\langle x \rangle \cdot \langle y \rangle$: Update MAC at each individual step of {\tt A4.1}-{\tt A4.3} accordingly by {\tt B1}-{\tt B3}.

\end{enumerate}
The additions and multiplications of $\llangle  x \rrangle$ and $\llangle  y \rrangle$ follow {\tt A1}-{\tt A4} and the MACs will be updated accordingly by {\tt B1}-{\tt B4}. 

To verify the computation of a function, it only requires to check the MACs of the revealed values and the final outcome, which can be checked all efficiently together in a batch at the final stage by a technique called ``random linear combination''.

\subsection{Pre-processing Phase}

In the pre-processing phase, all parties need to prepare a collection of triplets $(\langle  a \rangle, \langle b \rangle, \langle  c \rangle)$ where $c=a\cdot b$, each for a required multiplication operation.
Assume that the parties hold secretly shared numbers $a = \sum_{i=1}^N a_i$ and $b = \sum_{i=1}^N b_i$ (which has been generated by local random generation). Note that $a \cdot b= \sum_{i=1}^N a_i b_i + \sum_{i=1}^N \sum_{j=i"i\ne j}^N a_i b_j$. $a_i b_i$ can be computed locally. To distribute $a_i b_j$, one can use partial homomorphic cryptosystems, with encryption function ${\tt Enc}[\cdot]$ and decryption function ${\tt Dec}[\cdot]$ using party $i$'s public and private $(K_i^{\tt p}, K_i^{\tt p})$. First, party $i$ sends ${\tt Enc}_{K_i^{\tt p}}[a_i]$ to party $j$, who responds by $C_i = b_j {\tt Enc}_{K_i^{\tt p}}[a_i] - {\tt Enc}_{K_i^{\tt p}}[\tilde{c}_j]$, where $\tilde{c}_j$ is a random share generated by party $j$ and is encrypted by party $i$'s public key $K_i^{\tt s}$. Then party $i$ can obtain $\tilde{c}_j = {\tt Dec}_{K_i^{\tt s}}[C_j]$. Hence, $a_i b_j = \tilde{c}_i + \tilde{c}_j$, which are secret shares $a_i b_j$. The above generation assumes honest parties. To prevent cheating by dishonest parties, one would need to use proper zero-knowledge proofs before secret sharing \cite{cramer2015secure, dklpss13spdz}.

To generate a random mask $\llangle  r^i \rrangle$, each party $j$ needs to generate a random share $r^i_j$ locally. Then the parties follow the similar procedure of triplet generation to compute the secretly shared product $\langle  \gamma(r^i) \rangle$, where $\gamma(r^i) = \widetilde{\alpha} r^i$.

\subsection{Output and Validation Phase}
We describe random linear combination for batch checking. To check the MACs of a number of secretly shared numbers $\llangle  x^1 \rrangle, ..., \llangle  x^m \rrangle$ in a batch, first generate a set of random $({\tt r}^1, ..., {\tt r}^m)$. then reveal $\llangle  x^1 \rrangle, ..., \llangle  x^m \rrangle$. Each party $i$ computes $\sum_{j=1}^m {\tt r}^j (\widetilde{\alpha}_i - x^j \cdot \gamma(x^j)_i)$ and reveals it. All parties check whether $\sum_{i=1}^N \sum_{j=1}^m {\tt r}^j (\widetilde{\alpha}_i - x^j \cdot \gamma(x^j)_i) \overset{?}{=} 0$ for consistency in a batch checking.

\subsection{SPDZ Protocol}

We summarize the SPDZ protocol as follows:
\begin{enumerate}

\item {\em Pre-processing Phase}: In this phase, a collection of shared random numbers will be constructed that can be used to mask the private input numbers. For each private input number of party $i$, there is a shared random number $\llangle r^i \rrangle$, where $r^i$ is revealed to party $i$ only, but not to other parties. All parties also prepare a collection of triplets $(\llangle  a \rrangle, \llangle b \rrangle, \llangle  c \rrangle)$ where $c=a\cdot b$, each for a required multiplication operation.

\item {\em Online Phase}: To secretly shares a private input number $x^i$ using $\llangle r^i \rrangle$, without revealing $x^i$, it proceeds as follows:
\begin{enumerate}

\item[1)] Party $i$ computes and reveals $z^i = x^i - r^i$ to all parties.

\item[2)] Every party sets $\llangle x^i \rrangle \leftarrow z^i + \llangle r^i \rrangle$.

\end{enumerate}
\smallskip

To compute an arithmetic circuit, implement the required additions or multiplications by {\tt A1}-{\tt A4} and the MACs are updated accordingly by {\tt B1}-{\tt B4}. 

\item {\em Output and Validation Phase}: All MACs will be checked for all revealed numbers and the final output value. It can check all in a batch using random linear combination. If there is any inconsistency in the MACs, then abort.

\end{enumerate}

Note that SPDZ cannot guarantee abort with fairness -- dishonest parties may learn some partial values, even when the protocol aborts. However, this is a fundamental problem for any multi-party computation protocol with a dishonest majority, where dishonest parties are not identifiable when the computation is aborted.

%% file: append0.tex
\section{Additional Protocols}

\subsection{Basic Protocols with SPDZ} \label{sec:protocols}

In this section, we present several common protocols based on SPDZ as sub-routines:

\begin{enumerate}

\item $\Pi_{\rm RandBit}$ is a protocol that generates a secretly shared random bit $\llangle b \rrangle \in \{0, 1\}$ via SPDZ. 
 
 \item  $\Pi_{\rm RandPos}$ is a protocol that generates a secretly shared random positive number $\llangle R \rrangle > 0$ via SPDZ.
 
\item $\Pi_{\min}[\llangle y \rrangle,\llangle z \rrangle]$ is a protocol that computes, via SPDZ, secretly shared output $\llangle x \rrangle$ for given secretly shared values $\llangle y \rrangle,\llangle z \rrangle$, such that $x \leftarrow \min\{y, z\}$. 

\item $\Pi_{<}[\llangle x \rrangle,\llangle y \rrangle]$ is a protocol that compares, via SPDZ, two secretly shared values $\llangle x \rrangle$ and $\llangle y \rrangle$ and outputs 1 if $\llangle x \rrangle-\llangle y \rrangle<0$, and 0 otherwise, without revealing other information about $x, y$.

\item $\Pi_{=}[\llangle x \rrangle,\llangle y \rrangle]$ is a protocol that compares, via SPDZ, two secretly shared values $\llangle x \rrangle$ and $\llangle y \rrangle$ and outputs 1 if $\llangle x \rrangle-\llangle y \rrangle=0$, and 0 otherwise, without revealing other information about $x, y$.

\end{enumerate}

\begin{algorithm}[tbh!] 
	\caption{$\Pi_{\tt RandBit}$: {\em Generate a secretly shared random bit $\llangle b \rrangle \in \{0, 1\}$ via SPDZ, without revealing $b$}} \label{alg:randbit}
	{\scriptsize
	\begin{algorithmic}[1]		
		\Ensure $\llangle b \rrangle$
		\For{$i \in X$}
			\State User $i$ generates a random local bit $b_i \overset{\$}{\leftarrow} \{0, 1\}$
			\State Secretly share $b_i$ as $\llangle b_i \rrangle$ to all users
			\State Announce ${\tt Cm}(b_i)$ to all users
			\State Apply $\Pi_{\tt dzkpCm}[{\tt Cm}(b_i),\llangle b_i \rrangle]$ to prove the knowledge of $b_i$
			\State \mbox{User $i$ provides ${\tt nzkpMbs}[\{0,1\}, {\tt Cm}(b_i);b_i]$ to prove $\llangle b_i\rrangle$$\in$$\{0,1\}$} 
		\EndFor
		\State \mbox{Compute $\llangle b \rrangle \leftarrow \bigotimes_{i\in X} \llangle b_i \rrangle$ via SPDZ, where $\otimes$ is XOR operator:}
		$$
		\llangle b_{i} \rrangle \otimes \llangle b_{i'} \rrangle \triangleq 1 - \big(1 -  \llangle b_{i} \rrangle \cdot (1-\llangle b_{i'} \rrangle)\big) \cdot \big( 1- (1-\llangle b_{i}\rrangle) \cdot \llangle b_{i'}\rrangle \big)
		$$
		\State \Return $\llangle b \rrangle$
	\end{algorithmic}	
}	
\end{algorithm}

\begin{algorithm}[tbh!] 
	\caption{$\Pi_{\tt RandPos}$: {\em Generate a secretly shared random positive number $\llangle R \rrangle >0$ via SPDZ, without revealing $R$}} \label{alg:randpos}
	{\scriptsize
	\begin{algorithmic}[1]		
		\Ensure $\llangle R \rrangle$
		\For{$i \in X$}
			\State User $i$ generates a random local positive number $R_i \overset{\$}{\leftarrow} {\mathbb Z}_q$
			\State Secretly share $R_i$ as $\llangle R_i \rrangle$ to all users
			\State Announce ${\tt Cm}(R_i)$ to all users
			\State Apply $\Pi_{\tt dzkpCm}[{\tt Cm}(R_i),\llangle R_i \rrangle]$ to prove the knowledge of $R_i$
			\State User $i$ provides ${\tt nzkpNN}[{\tt Cm}(R_i);R_i]$ to prove $\llangle R_i \rrangle \ge 0$
		\EndFor	
		\State Compute $\llangle R \rrangle \leftarrow \sum_{i\in X} \llangle R_i \rrangle$ via SPDZ
		\State \Return $\llangle R \rrangle$ + 1
	\end{algorithmic}	
}	
\end{algorithm}

\begin{algorithm}[tbh!] 
	\caption{$\Pi_{\tt Min}$: {\em Compute $\llangle z \rrangle \leftarrow \min\{\llangle x \rrangle, \llangle y \rrangle\}$ via SPDZ, in a privacy-preserving manner without revealing $(z, x, y)$}} \label{alg:min}
	{\scriptsize
	\begin{algorithmic}[1]
		\Require $\llangle x \rrangle, \llangle y \rrangle$ (already secretly shared among users)
		\Ensure $\llangle z \rrangle$		
		\State Generate a secretly shared random bit $\llangle b \rrangle \leftarrow \Pi_{\tt RandBit}$			
		\State Randomly shuffle $(x, y)$ based on $b$ by the following via SPDZ:
		\begin{align}
		\llangle u \rrangle \leftarrow \llangle x \rrangle + \llangle b \rrangle\cdot(\llangle y \rrangle - \llangle x \rrangle) \notag\\
		\llangle v \rrangle \leftarrow \llangle y \rrangle + \llangle b \rrangle\cdot(\llangle x \rrangle-\llangle y \rrangle) \notag
		\end{align}
		\State \mbox{Generate a secretly shared random positive number $\llangle R \rrangle$$\leftarrow$$\Pi_{\tt RandPos}$}
		\State Compute $\llangle w \rrangle \leftarrow \llangle R \rrangle \cdot (\llangle u \rrangle - \llangle v \rrangle)$ via SPDZ
		\State Open $\llangle w \rrangle$ and its MAC to all users		
		\If{Checking MAC failed}
		\State Abort
		\EndIf
		\If{$w \ge 0$}
		\State Set $\llangle z \rrangle \leftarrow \llangle v \rrangle$ 
		\Else
		\State Set $\llangle z \rrangle \leftarrow \llangle u \rrangle$ 
		\EndIf
		\State \Return $\llangle z \rrangle$
	\end{algorithmic}		
}	
\end{algorithm}

\begin{algorithm}[tbh!] 
	\caption{$\Pi_{<}$: {\em Output 1 for given $\llangle x \rrangle, \llangle y \rrangle$, if $\llangle x \rrangle > \llangle y \rrangle$ via SPDZ, in a privacy-preserving manner without revealing $(x, y)$}} \label{alg:less}
	{\scriptsize
	\begin{algorithmic}[1]
		\Require $\llangle x \rrangle, \llangle y \rrangle$ (already secretly shared among users)
		\Ensure 1, if $\llangle x \rrangle > \llangle y \rrangle$. Otherwise, 0
		\State Generate a secretly shared random positive number $\llangle R \rrangle \leftarrow \Pi_{\tt RandPos}$			
		\State Compute $\llangle w \rrangle \leftarrow \llangle R \rrangle \cdot (\llangle x \rrangle - \llangle y \rrangle)$ via SPDZ
		\State Open $\llangle w \rrangle$ and its MAC to all users		
		\If{Checking MAC failed}
		\State Abort
		\EndIf	
		\If{$w > 0$}
		\State \Return 1
		\Else
		\State \Return 0
		\EndIf
	\end{algorithmic}		
}	
\end{algorithm}

\begin{algorithm}[tbh!] 
	\caption{$\Pi_{=}$: {\em Output 1 for given $\llangle x \rrangle, \llangle y \rrangle$, if $\llangle x \rrangle = \llangle y \rrangle$ via SPDZ, in a privacy-preserving manner without revealing $(x, y)$}} \label{alg:=}
	{\scriptsize
	\begin{algorithmic}[1]
		\Require $\llangle x \rrangle, \llangle y \rrangle$ (already secretly shared among users)
		\Ensure 1, if $\llangle x \rrangle = \llangle y \rrangle$. Otherwise, 0
		\State Generate a secretly shared random positive number $\llangle R \rrangle \leftarrow \Pi_{\tt RandPos}$			
		\State Compute $\llangle w \rrangle \leftarrow \llangle R \rrangle \cdot (\llangle x \rrangle - \llangle y \rrangle)$ via SPDZ
		\State Open $\llangle w \rrangle$ and its MAC to all users		
		\If{Checking MAC failed}
		\State Abort
		\EndIf
		\If{$w = 0$}
		\State \Return 1
		\Else
		\State \Return 0
		\EndIf
	\end{algorithmic}		
}	
\end{algorithm}

\subsection{\mbox{Privacy-preserving Payments on Blockchain}}\label{sec:payments}

On Ethereum, one can create tokens on the ledger to represent certain digital assets. Our mutual compensation payment system is implemented by ERC20 tokens \cite{erc20}. To make payment among each other, users are required to purchase tokens that will be subsequently transferred to each other and redeemed. By default, the transaction records on the ledger are completely visible to the public. In this section, we incorporate privacy protection to hide the transaction records on the ledger. As in other privacy-preserving blockchain platforms (e.g., Zether \cite{bunz20zether}), we conceal the balances and transaction values in the ledger by the respective cryptographic commitments instead of plaintext values. For example, ${\tt Cm}({\tt Bal}({\tt ad}_i))$ will be recorded as the balance for account ${\tt ad}_i$ on the ledger.

To initiate a transaction of tokens from ${\tt ad}_i$ to ${\tt ad}_{i'}$ with transaction value ${\tt val}$, a user submits a transaction request to the blockchain: ${\tt tx} = ({\tt ad}_i, {\tt ad}_{i'}, {\tt val})$, along with a signature ${\tt sign}_{K^{\tt s}_i}({\tt tx})$ using the private key $K^{\tt s}_i$ associated with ${\tt ad}_i$. To pay mutual compensations among multiple users, the users submit a {\em multi-transaction}, denoted by ${\tt mtx} = ({\tt ad}_i, {\tt ad}_{i'}, {\tt val}_i)_{i=1}^N$, which will be executed, only if ${\tt Bal}({\tt ad}_i) \ge {\tt val}_i$ for all $i$ and multi-signature ${\tt sign}_{(K^{\tt s}_i)_{i=1}^N}({\tt mtx})$ is present. To hide transfer amounts, a multi-transaction can be concealed as ${\tt mtx}=({\tt ad}_i, {\tt ad}_{i'}, {\tt Cm}({\tt val}_i))_{i=1}^N$. Since Pedersen commitment satisfies homomorphic property, the concealed transaction value can be added to the concealed balance as follows: $${\tt Cm}({\tt Bal}({\tt ad}_i))\leftarrow{\tt Cm}({\tt Bal}({\tt ad}_i))\cdot{\tt Cm}({\tt val}_i)$$
However, the transaction may be invalid, when ${\tt Bal}({\tt ad}_i) \le {\tt val}_i$. Hence, each user must provide ${\tt nzkpNN}[{\tt Bal}({\tt ad}_i) - {\tt val}_i\ge 0]$ along with each transaction request to prove the non-negativity of the resultant balance. Otherwise, the transaction request will be denied.

In Stage 4, each user $i$ pays the net mutual compensation amount $\phi_i$ (which may be negative, if the user receives compensation). Note that it is evident that $\sum_{i \in X} \phi_i = 0$ (namely, the sum of net mutual compensations should be zero). Hence, it suffices to consider the scenario that all users transfer the payments to a dummy address ${\tt ad}_{\tt dummy}$ with zero total transaction value, with each transferring a transaction value of ${\tt Cm}(\phi_i)$ in a commitment. To prove the validity of the multi-transaction, the users need to provide a ZKP of summation, such that the summation of all transferred amounts equals zero, namely, $\prod_{i \in X} {\tt Cm}(\phi_i) = {\tt Cm}(0)$.
This can be constructed by $\Sigma$-protocol. However, we need a distributed version of ZKP of summation, as it can be validated by all users. A distributed  ZKP of summation via SPDZ is described in $\Pi_{\tt dzkpsum}$ (Algorithm~\ref{alg:dzkpsum}).

\begin{algorithm}[tbh!] 
	\caption{$\Pi_{\tt dzkpSum}$: {\em Prove the knowledge of $y = \sum_{i=1}^n x_i$ in given commitments $\big( {\tt Cm}(x_1, {\tt r}_1)$,...,${\tt Cm}(x_n, {\tt r}_n) \big)$ and  public known value $y$, via SPDZ}} \label{alg:dzkpsum}
	{\scriptsize
	\begin{algorithmic}[1]		
		\Require $\big( {\tt Cm}(x_1, {\tt r}_1)$,...,${\tt Cm}(x_n, {\tt r}_n) \big)$ (known to the verifier and prover) 
		\Ensure {\sf Pass} or {\sf Fail}
		
		\State Each user $i$ announces ${\tt Cm}(x_i, {\tt r}_i)$ and secretly shares $\llangle {\tt r}_i \rrangle$ with all users
		
		\State Each user $i$ randomly generates ${\tt r}'_i \leftarrow {\mathbb Z}_p$ and secretly shares $\llangle {\tt r}'_i \rrangle$ before announcing ${\tt Cm}(0, {\tt r}'_{i})$ to all users
		
		\State All users compute $C' \leftarrow \prod_{i=1}^N {\tt Cm}(0, {\tt r}'_{i})$ and obtain a random challenge $\beta \leftarrow {\mathcal H}(C')$
		
		\State All users compute $\llangle z_{\tt r} \rrangle \leftarrow \sum_{i=1}^n \llangle {\tt r}'_i \rrangle + \beta \cdot \sum_{i=1}^n \llangle {\tt r}_i \rrangle$ via SPDZ
		
		\State Reveal $\llangle z_{\tt r} \rrangle$ and its MAC to all users

		\Statex \LeftComment{{\em All users check the following}}
		\If{$g^{\beta \cdot y}\cdot h^{z_{\tt r}} \overset{?}{=}  C'\cdot\prod_{i=1}^n {\tt Cm}(x_i, {\tt r}_i)^\beta$ and checking MAC passed} 
		
		\State \Return {\sf Pass} 

		\Else

		\State \Return {\sf Fail}
		
		\EndIf
		
		\Statex \LeftComment{{\em The ZKP of summation is ${\sf zkpSum}[\sum_{i=1}^n x_i = y] = \big\{ \big( {\tt Cm}(x_1, {\tt r}_1)$,...,${\tt Cm}(x_n, {\tt r}_n) \big); C', z_{\tt r} \big\}$}}		
		
	\end{algorithmic}	
	}
\end{algorithm}

Next, we describe the process of privacy-preserving payments of net mutual payments as follows:
\begin{enumerate}

\item Each user $i$ retains secretly-shared $\big(\llangle \phi_i \rrangle\big)_{i \in X}$ in Stage 3, and announces its commitment ${\tt Cm}(\phi_i, {\tt r}_i)$ to all users, and then secretly shares $\llangle {\tt r}_i \rrangle$ via SPDZ.

\item All users generate ZKP of summation by $\Pi_{\tt dzkpsum}$ to prove that $\sum_{i\in X} \phi_i  = 0$.

\item Each user $i$ generates ${\tt nzkpNN}[{\tt Bal}({\tt ad}_i) - \phi_i\ge 0]$ based on ${\tt Cm}(\phi_i, {\tt r}_i)$.

\item The users submit a multi-transaction request 
\[
{\tt mtx} = \big({\tt ad}_i, {\tt ad}_{\tt dummy}, {\tt Cm}(\phi_i, {\tt r}_i)\big)_{i\in X}
\]
to the blockchain ledger, along with the following ZKPs:
\[
\qquad \ \ {\tt nzkpSum}[\sum_{i\in X} \phi_i  = 0 ] \mbox{\ and\ } {\tt nzkpNN}[{\tt Bal}({\tt ad}_i) - \phi_i\ge 0]_{i\in X}
\]

\item The blockchain ledger verifies ${\tt nzkpSum}$ and ${\tt nzkpNN}$ before recording the transaction.

\end{enumerate}

%% file: paper.bbl

\begin{thebibliography}{42}

%%% the Web address.

\ifx \showCODEN    \undefined \def \showCODEN     #1{\unskip}     \fi
\ifx \showDOI      \undefined \def \showDOI       #1{#1}\fi
\ifx \showISBNx    \undefined \def \showISBNx     #1{\unskip}     \fi
\ifx \showISBNxiii \undefined \def \showISBNxiii  #1{\unskip}     \fi
\ifx \showISSN     \undefined \def \showISSN      #1{\unskip}     \fi
\ifx \showLCCN     \undefined \def \showLCCN      #1{\unskip}     \fi
\ifx \shownote     \undefined \def \shownote      #1{#1}          \fi
\ifx \showarticletitle \undefined \def \showarticletitle #1{#1}   \fi
\ifx \showURL      \undefined \def \showURL       {\relax}        \fi
% invisible to TeX
\providecommand\bibfield[2]{#2}
\providecommand\bibinfo[2]{#2}
\providecommand\natexlab[1]{#1}
\providecommand\showeprint[2][]{arXiv:#2}

\bibitem[\protect\citeauthoryear{Albers and Quedenfeld}{Albers and Quedenfeld}{2018}]%
        {disc}
\bibfield{author}{\bibinfo{person}{Susanne Albers} {and} \bibinfo{person}{Jens Quedenfeld}.} \bibinfo{year}{2018}\natexlab{}.
\newblock \showarticletitle{Optimal algorithms for right-sizing data centers}.
\newblock \bibinfo{journal}{\emph{SPAA}} (\bibinfo{year}{2018}).
\newblock


\bibitem[\protect\citeauthoryear{Allan, Linial, and Saks}{Allan et~al\mbox{.}}{1992}]%
        {mts}
\bibfield{author}{\bibinfo{person}{Borodin Allan}, \bibinfo{person}{Nathan Linial}, {and} \bibinfo{person}{Michael~E. Saks}.} \bibinfo{year}{1992}\natexlab{}.
\newblock \showarticletitle{An optimal on-line algorithm for metrical task system}.
\newblock \bibinfo{journal}{\emph{JACM}} (\bibinfo{year}{1992}).
\newblock


\bibitem[\protect\citeauthoryear{Andoni, Robu, Flynn, Abram, Geach, Jenkins, McCallum, and Peacock}{Andoni et~al\mbox{.}}{2019}]%
        {a19survey}
\bibfield{author}{\bibinfo{person}{Merlinda Andoni}, \bibinfo{person}{Valentin Robu}, \bibinfo{person}{David Flynn}, \bibinfo{person}{Simone Abram}, \bibinfo{person}{Dale Geach}, \bibinfo{person}{David~P. Jenkins}, \bibinfo{person}{Peter McCallum}, {and} \bibinfo{person}{Andrew Peacock}.} \bibinfo{year}{2019}\natexlab{}.
\newblock \showarticletitle{Blockchain technology in the energy sector: A systematic review of challenges and opportunities}.
\newblock \bibinfo{journal}{\emph{Renewable and Sustainable Energy Reviews}}  \bibinfo{volume}{100} (\bibinfo{year}{2019}), \bibinfo{pages}{143--174}.
\newblock


\bibitem[\protect\citeauthoryear{Bansal, Gupta, Krishnaswamy, Pruhs, Schewior, and Stein}{Bansal et~al\mbox{.}}{2015}]%
        {oco}
\bibfield{author}{\bibinfo{person}{Nikhil Bansal}, \bibinfo{person}{Anupam Gupta}, \bibinfo{person}{Ravishankar Krishnaswamy}, \bibinfo{person}{Kirk Pruhs}, \bibinfo{person}{Kevin Schewior}, {and} \bibinfo{person}{Cliff Stein}.} \bibinfo{year}{2015}\natexlab{}.
\newblock \showarticletitle{A 2-competitive algorithm for online convex optimization with switching costs}.
\newblock \bibinfo{journal}{\emph{APPROX}} (\bibinfo{year}{2015}).
\newblock


\bibitem[\protect\citeauthoryear{Ben-Or, Goldwasser, and Wigderson}{Ben-Or et~al\mbox{.}}{1988}]%
        {BGW99}
\bibfield{author}{\bibinfo{person}{Michael Ben-Or}, \bibinfo{person}{Shafi Goldwasser}, {and} \bibinfo{person}{Avi Wigderson}.} \bibinfo{year}{1988}\natexlab{}.
\newblock \showarticletitle{Completeness Theorems for Non-Cryptographic Fault-Tolerant Distributed Computation}. In \bibinfo{booktitle}{\emph{Annual ACM Symposium on Theory of Computing (STOC)}}.
\newblock


\bibitem[\protect\citeauthoryear{Ben-Sasson, Chiesa, Garman, Green, Miers, and Virza}{Ben-Sasson et~al\mbox{.}}{2014}]%
        {zcash}
\bibfield{author}{\bibinfo{person}{Eli Ben-Sasson}, \bibinfo{person}{Alessandro Chiesa}, \bibinfo{person}{Christina Garman}, \bibinfo{person}{Matthew Green}, \bibinfo{person}{Ian Miers}, {and} \bibinfo{person}{Eran Tromerand~Madars Virza}.} \bibinfo{year}{2014}\natexlab{}.
\newblock \showarticletitle{Zerocash: Decentralized Anonymous Payments from Bitcoin}. In \bibinfo{booktitle}{\emph{IEEE Symposium on Security and Privacy}}.
\newblock


\bibitem[\protect\citeauthoryear{Borodin and El-Yaniv}{Borodin and El-Yaniv}{2005}]%
        {BEY05online}
\bibfield{author}{\bibinfo{person}{Allan Borodin} {and} \bibinfo{person}{Ran El-Yaniv}.} \bibinfo{year}{2005}\natexlab{}.
\newblock \bibinfo{booktitle}{\emph{Online Computation and Competitive Analysis}}.
\newblock \bibinfo{publisher}{Cambridge University Press}.
\newblock


\bibitem[\protect\citeauthoryear{Buchanan}{Buchanan}{2017}]%
        {crytobk}
\bibfield{author}{\bibinfo{person}{William~J. Buchanan}.} \bibinfo{year}{2017}\natexlab{}.
\newblock \bibinfo{booktitle}{\emph{Cryptography}}.
\newblock \bibinfo{publisher}{River Publishers}.
\newblock


\bibitem[\protect\citeauthoryear{Bunz, Agrawal, Zamani, and Boneh}{Bunz et~al\mbox{.}}{2020}]%
        {bunz20zether}
\bibfield{author}{\bibinfo{person}{Benedikt Bunz}, \bibinfo{person}{Shashank Agrawal}, \bibinfo{person}{Mahdi Zamani}, {and} \bibinfo{person}{Dan Boneh}.} \bibinfo{year}{2020}\natexlab{}.
\newblock \showarticletitle{Zether: Towards Privacy in a Smart Contract World}. In \bibinfo{booktitle}{\emph{Financial Cryptography and Data Security (FC)}}.
\newblock


\bibitem[\protect\citeauthoryear{Chau and Elbassioni}{Chau and Elbassioni}{2018}]%
        {CE17sharing}
\bibfield{author}{\bibinfo{person}{Chi-Kin Chau} {and} \bibinfo{person}{Khaled Elbassioni}.} \bibinfo{year}{2018}\natexlab{}.
\newblock \showarticletitle{Quantifying Inefficiency of Fair Cost-Sharing Mechanisms for Sharing Economy}.
\newblock \bibinfo{journal}{\emph{IEEE Trans. Control of Network System}}  \bibinfo{volume}{5} (\bibinfo{date}{Dec} \bibinfo{year}{2018}), \bibinfo{pages}{1809--1818}.
\newblock
Issue 4.
\newblock
\shownote{https://arxiv.org/abs/1511.05270.}


\bibitem[\protect\citeauthoryear{Chau, Elbassioni, and Zhou}{Chau et~al\mbox{.}}{2022}]%
        {CE20sharing}
\bibfield{author}{\bibinfo{person}{Sid Chi-Kin Chau}, \bibinfo{person}{Khaled Elbassioni}, {and} \bibinfo{person}{Yue Zhou}.} \bibinfo{year}{2022}\natexlab{}.
\newblock \bibinfo{booktitle}{\emph{Approximately Socially-Optimal Decentralized Coalition Formation with Application to P2P Energy Sharing}}.
\newblock \bibinfo{type}{{T}echnical {R}eport}.
\newblock
\newblock
\shownote{https://arxiv.org/abs/2009.08632.}


\bibitem[\protect\citeauthoryear{Chau, Xu, Bow, and Elbassioni}{Chau et~al\mbox{.}}{2019}]%
        {p2p19}
\bibfield{author}{\bibinfo{person}{Sid Chi-Kin Chau}, \bibinfo{person}{Jiajia Xu}, \bibinfo{person}{Wilson Bow}, {and} \bibinfo{person}{Khaled Elbassioni}.} \bibinfo{year}{2019}\natexlab{}.
\newblock \showarticletitle{Peer-to-Peer Energy Sharing: Effective Cost-Sharing Mechanisms and Social Efficiency}. In \bibinfo{booktitle}{\emph{ACM Intl. Conf. on Future Energy Systems (e-Energy)}}.
\newblock


\bibitem[\protect\citeauthoryear{Chau and Zhou}{Chau and Zhou}{2022}]%
        {CZ22energyplan}
\bibfield{author}{\bibinfo{person}{Sid Chi-Kin Chau} {and} \bibinfo{person}{Yue Zhou}.} \bibinfo{year}{2022}\natexlab{}.
\newblock \showarticletitle{{Blockchain-Enabled Decentralized Privacy-Preserving Group Purchasing for Retail Energy Plans}}. In \bibinfo{booktitle}{\emph{Proc. ACM Intl. Conf. on Future Energy Systems (e-Energy)}}.
\newblock


\bibitem[\protect\citeauthoryear{Chen and Roma}{Chen and Roma}{2011}]%
        {CR11group}
\bibfield{author}{\bibinfo{person}{Rachel~R. Chen} {and} \bibinfo{person}{Paolo Roma}.} \bibinfo{year}{2011}\natexlab{}.
\newblock \showarticletitle{Group buying of competing retailers}.
\newblock \bibinfo{journal}{\emph{Production and Operations Management}} \bibinfo{volume}{20}, \bibinfo{number}{2} (\bibinfo{year}{2011}), \bibinfo{pages}{181--197}.
\newblock


\bibitem[\protect\citeauthoryear{Chen, Pastro, and Raykova}{Chen et~al\mbox{.}}{2018}]%
        {cpr18spdz}
\bibfield{author}{\bibinfo{person}{Valerie Chen}, \bibinfo{person}{Valerio Pastro}, {and} \bibinfo{person}{Mariana Raykova}.} \bibinfo{year}{2018}\natexlab{}.
\newblock \showarticletitle{Secure Computation for Machine Learning With SPDZ}. In \bibinfo{booktitle}{\emph{Annual Conference on Neural Information Processing Systems (NeurIPS)}}.
\newblock


\bibitem[\protect\citeauthoryear{ChoiceEnergy}{ChoiceEnergy}{2021}]%
        {choiceenergy}
\bibfield{author}{\bibinfo{person}{ChoiceEnergy}.} \bibinfo{year}{2021}\natexlab{}.
\newblock \bibinfo{title}{Cut energy costs with the bulk-buying power of a group tender}.
\newblock \bibinfo{howpublished}{\url{https://www.choiceenergy.com.au/group-energy-tenders}}.   (\bibinfo{year}{2021}).
\newblock


\bibitem[\protect\citeauthoryear{Cramer, Damg{\aa}rd, and Nielsen}{Cramer et~al\mbox{.}}{2001}]%
        {cramer2001multiparty}
\bibfield{author}{\bibinfo{person}{Ronald Cramer}, \bibinfo{person}{Ivan Damg{\aa}rd}, {and} \bibinfo{person}{Jesper~B Nielsen}.} \bibinfo{year}{2001}\natexlab{}.
\newblock \showarticletitle{Multiparty computation from threshold homomorphic encryption}. In \bibinfo{booktitle}{\emph{Intl. conference on the theory and applications of cryptographic techniques}}.
\newblock


\bibitem[\protect\citeauthoryear{Cramer, Damg{\aa}rd, and Nielsen}{Cramer et~al\mbox{.}}{2015}]%
        {cramer2015secure}
\bibfield{author}{\bibinfo{person}{Ronald Cramer}, \bibinfo{person}{Ivan~Bjerre Damg{\aa}rd}, {and} \bibinfo{person}{Jesper~Buus Nielsen}.} \bibinfo{year}{2015}\natexlab{}.
\newblock \bibinfo{booktitle}{\emph{Secure Multiparty Computation and Secret Sharing}}.
\newblock \bibinfo{publisher}{Cambridge University Press}.
\newblock
\newblock
\shownote{Cambridge Books Online.}


\bibitem[\protect\citeauthoryear{Damg{\aa}rd, Keller, Larraia, Pastro, Scholl, and Smart}{Damg{\aa}rd et~al\mbox{.}}{2013}]%
        {dklpss13spdz}
\bibfield{author}{\bibinfo{person}{Ivan Damg{\aa}rd}, \bibinfo{person}{Marcel Keller}, \bibinfo{person}{Enrique Larraia}, \bibinfo{person}{Valerio Pastro}, \bibinfo{person}{Peter Scholl}, {and} \bibinfo{person}{Nigel~P. Smart}.} \bibinfo{year}{2013}\natexlab{}.
\newblock \showarticletitle{Practical Covertly Secure MPC for Dishonest Majority - or: Breaking the SPDZ Limits}. In \bibinfo{booktitle}{\emph{European Symposium on Research in Computer Security (ESORICS)}}.
\newblock


\bibitem[\protect\citeauthoryear{Du and Atallah}{Du and Atallah}{2001}]%
        {du2001secure}
\bibfield{author}{\bibinfo{person}{Wenliang Du} {and} \bibinfo{person}{Mikhail~J Atallah}.} \bibinfo{year}{2001}\natexlab{}.
\newblock \showarticletitle{Secure multi-party computation problems and their applications: a review and open problems}. In \bibinfo{booktitle}{\emph{the Workshop on New Security Paradigms}}.
\newblock


\bibitem[\protect\citeauthoryear{Dwork, McSherry, Nissim, and Smith}{Dwork et~al\mbox{.}}{2006}]%
        {dwork2006calibrating}
\bibfield{author}{\bibinfo{person}{Cynthia Dwork}, \bibinfo{person}{Frank McSherry}, \bibinfo{person}{Kobbi Nissim}, {and} \bibinfo{person}{Adam Smith}.} \bibinfo{year}{2006}\natexlab{}.
\newblock \showarticletitle{Calibrating noise to sensitivity in private data analysis}. In \bibinfo{booktitle}{\emph{Theory of cryptography conference}}. Springer.
\newblock


\bibitem[\protect\citeauthoryear{Ethereum.org}{Ethereum.org}{2022}]%
        {erc20}
\bibfield{author}{\bibinfo{person}{Ethereum.org}.} \bibinfo{year}{2022}\natexlab{}.
\newblock \bibinfo{title}{ERC-20 Token Standar}.
\newblock \bibinfo{howpublished}{\url{https://ethereum.org/en/developers/docs/standards/tokens/erc-20/}}.   (\bibinfo{year}{2022}).
\newblock


\bibitem[\protect\citeauthoryear{Gas}{Gas}{2021}]%
        {eastcoast}
\bibfield{author}{\bibinfo{person}{East Coast Power~\& Gas}.} \bibinfo{year}{2021}\natexlab{}.
\newblock \bibinfo{title}{General Temrs \& Conditions -- New York}.
\newblock \bibinfo{howpublished}{\url{http://www.ecpg.com/terms}}.   (\bibinfo{year}{2021}).
\newblock


\bibitem[\protect\citeauthoryear{Goldreich}{Goldreich}{1998}]%
        {goldreich1998secure}
\bibfield{author}{\bibinfo{person}{Oded Goldreich}.} \bibinfo{year}{1998}\natexlab{}.
\newblock \showarticletitle{Secure multi-party computation}.
\newblock \bibinfo{journal}{\emph{Manuscript. Preliminary version}}  \bibinfo{volume}{78} (\bibinfo{year}{1998}).
\newblock


\bibitem[\protect\citeauthoryear{Gorenflo, Golab, and Keshav}{Gorenflo et~al\mbox{.}}{2019}]%
        {ggk19block}
\bibfield{author}{\bibinfo{person}{Christian Gorenflo}, \bibinfo{person}{Lukasz Golab}, {and} \bibinfo{person}{Srinivasan Keshav}.} \bibinfo{year}{2019}\natexlab{}.
\newblock \showarticletitle{Using a Blockchain to Mitigate Trust in Electric Vehicle Charging}. In \bibinfo{booktitle}{\emph{ACM Intl. Conf. on Future Energy Systems (e-Energy)}}.
\newblock


\bibitem[\protect\citeauthoryear{Hajiesmaili, Chau, Chen, and Huang}{Hajiesmaili et~al\mbox{.}}{2016}]%
        {rchase}
\bibfield{author}{\bibinfo{person}{Mohammad~H. Hajiesmaili}, \bibinfo{person}{Chi-Kin Chau}, \bibinfo{person}{Minghua Chen}, {and} \bibinfo{person}{Longbu Huang}.} \bibinfo{year}{2016}\natexlab{}.
\newblock \showarticletitle{Online Microgrid Energy Generation Scheduling Revisited: The Benefits of Randomization and Interval Prediction}.
\newblock \bibinfo{journal}{\emph{ACM e-Energy}} (\bibinfo{year}{2016}).
\newblock


\bibitem[\protect\citeauthoryear{Hazay and Lindell}{Hazay and Lindell}{2010}]%
        {HazayL10}
\bibfield{author}{\bibinfo{person}{Carmit Hazay} {and} \bibinfo{person}{Yehuda Lindell}.} \bibinfo{year}{2010}\natexlab{}.
\newblock \bibinfo{booktitle}{\emph{Efficient Secure Two-Party Protocols - Techniques and Constructions}}.
\newblock \bibinfo{publisher}{Springer}.
\newblock


\bibitem[\protect\citeauthoryear{Knirsch, Brunner, Unterweger, and Engel}{Knirsch et~al\mbox{.}}{2020}]%
        {kbue20gecko}
\bibfield{author}{\bibinfo{person}{Fabian Knirsch}, \bibinfo{person}{Clemens Brunner}, \bibinfo{person}{Andreas Unterweger}, {and} \bibinfo{person}{Dominik Engel}.} \bibinfo{year}{2020}\natexlab{}.
\newblock \showarticletitle{Decentralized and permission-less green energy certificates with GECKO}.
\newblock \bibinfo{journal}{\emph{Energy Informatics}} \bibinfo{volume}{3}, \bibinfo{number}{2} (\bibinfo{year}{2020}).
\newblock


\bibitem[\protect\citeauthoryear{Lin, Wierman, Andrew, and Thereska}{Lin et~al\mbox{.}}{2013}]%
        {lcp}
\bibfield{author}{\bibinfo{person}{Minghong Lin}, \bibinfo{person}{Adam Wierman}, \bibinfo{person}{Lachlan L.~H. Andrew}, {and} \bibinfo{person}{Eno Thereska}.} \bibinfo{year}{2013}\natexlab{}.
\newblock \showarticletitle{Dynamic right-sizing for power-proportional data centers}.
\newblock \bibinfo{journal}{\emph{IEEE/ACM TON}} (\bibinfo{year}{2013}).
\newblock


\bibitem[\protect\citeauthoryear{Lu, Tu, Chau, Chen, and Lin}{Lu et~al\mbox{.}}{2013}]%
        {chase}
\bibfield{author}{\bibinfo{person}{Lian Lu}, \bibinfo{person}{Jinlong Tu}, \bibinfo{person}{Chi-Kin Chau}, \bibinfo{person}{Minghua Chen}, {and} \bibinfo{person}{Xiaojun Lin}.} \bibinfo{year}{2013}\natexlab{}.
\newblock \showarticletitle{Online energy generation scheduling for microgrids with intermittent energy sources and co-generation}.
\newblock \bibinfo{journal}{\emph{ACM SIGMETRICS}} (\bibinfo{year}{2013}).
\newblock


\bibitem[\protect\citeauthoryear{Lyu, Chau, Wang, and Zheng}{Lyu et~al\mbox{.}}{2020}]%
        {lcwz20privsharing}
\bibfield{author}{\bibinfo{person}{Lingjuan Lyu}, \bibinfo{person}{Sid Chi-Kin Chau}, \bibinfo{person}{Nan Wang}, {and} \bibinfo{person}{Yifeng Zheng}.} \bibinfo{year}{2020}\natexlab{}.
\newblock \showarticletitle{Cloud-based Privacy-Preserving Collaborative Consumption for Sharing Economy}.
\newblock \bibinfo{journal}{\emph{IEEE Trans. Cloud Computing}} (\bibinfo{year}{2020}).
\newblock


\bibitem[\protect\citeauthoryear{Menati, Chau, and Chen}{Menati et~al\mbox{.}}{2022}]%
        {pchase}
\bibfield{author}{\bibinfo{person}{Ali Menati}, \bibinfo{person}{Chi-Kin Chau}, {and} \bibinfo{person}{Minghua Chen}.} \bibinfo{year}{2022}\natexlab{}.
\newblock \showarticletitle{Competitive Prediction-Aware Online Algorithms for Energy Generation Scheduling in Microgrids}.
\newblock \bibinfo{journal}{\emph{ACM e-Energy}} (\bibinfo{year}{2022}).
\newblock


\bibitem[\protect\citeauthoryear{Mengelkamp, Garttner, Rock, Kessler, Orsini, and Weinhardt}{Mengelkamp et~al\mbox{.}}{2018}]%
        {m18brooklyn}
\bibfield{author}{\bibinfo{person}{Esther Mengelkamp}, \bibinfo{person}{Johannes Garttner}, \bibinfo{person}{Kerstin Rock}, \bibinfo{person}{Scott Kessler}, \bibinfo{person}{Lawrence Orsini}, {and} \bibinfo{person}{Christof Weinhardt}.} \bibinfo{year}{2018}\natexlab{}.
\newblock \showarticletitle{Designing microgrid energy markets: A case study: The Brooklyn Microgrid}.
\newblock \bibinfo{journal}{\emph{Applied Energy}}  \bibinfo{volume}{210} (\bibinfo{year}{2018}), \bibinfo{pages}{870--880}.
\newblock


\bibitem[\protect\citeauthoryear{Monero}{Monero}{2021}]%
        {monero}
\bibfield{author}{\bibinfo{person}{Monero}.} \bibinfo{year}{2021}\natexlab{}.
\newblock \bibinfo{howpublished}{\url{http://getmonero.org}}.   (\bibinfo{year}{2021}).
\newblock


\bibitem[\protect\citeauthoryear{Paper}{Paper}{2014}]%
        {ethereum}
\bibfield{author}{\bibinfo{person}{The Ethereum~Yellow Paper}.} \bibinfo{year}{2014}\natexlab{}.
\newblock \bibinfo{howpublished}{\url{https://ethereum.github.io/yellowpaper/paper.pdf}}.   (\bibinfo{year}{2014}).
\newblock


\bibitem[\protect\citeauthoryear{Radi, Lasla, Bakiras, and Mahmoud}{Radi et~al\mbox{.}}{2019}]%
        {RL19EV}
\bibfield{author}{\bibinfo{person}{Eman~Mohammed Radi}, \bibinfo{person}{Noureddine Lasla}, \bibinfo{person}{Spiridon Bakiras}, {and} \bibinfo{person}{Mohamed Mahmoud}.} \bibinfo{year}{2019}\natexlab{}.
\newblock \showarticletitle{Privacy-Preserving Electric Vehicle Charging for Peer-to-Peer Energy Trading Ecosystems}. In \bibinfo{booktitle}{\emph{IEEE ICC}}.
\newblock


\bibitem[\protect\citeauthoryear{Regulator}{Regulator}{2021}]%
        {energymadeeasy}
\bibfield{author}{\bibinfo{person}{Australian~Energy Regulator}.} \bibinfo{year}{2021}\natexlab{}.
\newblock \bibinfo{title}{Energy Made Easy}.
\newblock \bibinfo{howpublished}{\url{https://www.energymadeeasy.gov.au/}}.   (\bibinfo{year}{2021}).
\newblock


\bibitem[\protect\citeauthoryear{Wang, Chau, and Zhou}{Wang et~al\mbox{.}}{2021}]%
        {CWZ21blockchain}
\bibfield{author}{\bibinfo{person}{Nan Wang}, \bibinfo{person}{Sid Chi-Kin Chau}, {and} \bibinfo{person}{Yue Zhou}.} \bibinfo{year}{2021}\natexlab{}.
\newblock \showarticletitle{Privacy-Preserving Energy Storage Sharing with Blockchain}. In \bibinfo{booktitle}{\emph{ACM e-Energy 21'}}.
\newblock


\bibitem[\protect\citeauthoryear{Zhang, Chau, and Chen}{Zhang et~al\mbox{.}}{2019}]%
        {ZCC19energyplan}
\bibfield{author}{\bibinfo{person}{Jianing Zhang}, \bibinfo{person}{Sid Chi-Kin Chau}, {and} \bibinfo{person}{Minghua Chen}.} \bibinfo{year}{2019}\natexlab{}.
\newblock \showarticletitle{Stay or Switch: Competitive Online Algorithms for Energy Plan Selection in Energy Markets with Retail Choice}. In \bibinfo{booktitle}{\emph{ACM e-Energy 19'}}.
\newblock


\bibitem[\protect\citeauthoryear{Zhao, Jung, Wang, and Li}{Zhao et~al\mbox{.}}{2014}]%
        {ZJ14smeter}
\bibfield{author}{\bibinfo{person}{Jing Zhao}, \bibinfo{person}{Taeho Jung}, \bibinfo{person}{Yu Wang}, {and} \bibinfo{person}{Xiangyang Li}.} \bibinfo{year}{2014}\natexlab{}.
\newblock \showarticletitle{Achieving differential privacy of data disclosure in the smart grid}. In \bibinfo{booktitle}{\emph{IEEE Infocom}}.
\newblock


\bibitem[\protect\citeauthoryear{Zhou}{Zhou}{2017}]%
        {zhou17retail}
\bibfield{author}{\bibinfo{person}{Shengru Zhou}.} \bibinfo{year}{2017}\natexlab{}.
\newblock \bibinfo{booktitle}{\emph{An Introduction to Retail Electricity Choice in the United States}}.
\newblock \bibinfo{type}{{T}echnical {R}eport}. \bibinfo{institution}{NREL}.
\newblock


\bibitem[\protect\citeauthoryear{Zhou and Chau}{Zhou and Chau}{2021}]%
        {ZC21energyplan}
\bibfield{author}{\bibinfo{person}{Yue Zhou} {and} \bibinfo{person}{Sid Chi-Kin Chau}.} \bibinfo{year}{2021}\natexlab{}.
\newblock \showarticletitle{Sharing Economy Meets Energy Markets: Group Purchasing of Energy Plans in Retail Energy Markets}. In \bibinfo{booktitle}{\emph{ACM BuildSys 21'}}.
\newblock


\end{thebibliography}
